\documentclass[acmsmall]{acmart}

\setcopyright{cc}
\setcctype{by}
\acmDOI{10.1145/3763092}
\acmYear{2025}
\acmJournal{PACMPL}
\acmVolume{9}
\acmNumber{OOPSLA2}
\acmArticle{314}
\acmMonth{10}
\received{2025-03-25}
\received[accepted]{2025-08-12}

\usepackage{url}
\usepackage{wrapfig}
\usepackage{algorithmic}
\usepackage{graphicx}
\usepackage{subcaption}
\usepackage[export]{adjustbox}
\usepackage{textcomp}
\usepackage{xcolor}
\usepackage{makecell}
\usepackage{listings}
\usepackage{color}
\usepackage{multirow}
\usepackage{tabularx}
\usepackage{colortbl}
\usepackage{bm}
\usepackage{multicol}
\usepackage{algorithm}
\usepackage{algorithmic}
\usepackage{amsmath}
\usepackage{enumitem}
\usepackage{amsmath}
\usepackage{float}
\usepackage[absolute,overlay]{textpos}
\pdfmapfile{+helvetic.map}

\definecolor{darkgreen}{rgb}{0.0, 0.6, 0.0}
\definecolor{darkblue}{rgb}{0.0, 0.0, 0.6}

\newcommand{\byop}[0]{\textcolor{darkblue}{\textbf{\texttt{by}}}}
\newcommand{\systemname}[0]{MTP}

\lstdefinelanguage{yaml}{
  keywords={true,false,null,y,n},
  keywordstyle=\color{darkgray}\bfseries,
  sensitive=false,
  comment=[l]{\#},
  commentstyle=\color{purple}\ttfamily,
  stringstyle=\color{blue}\ttfamily,
  morestring=[b]',
  morestring=[b]"
}

\lstdefinelanguage{python}{
   keywords={sem, import, node, ignore, take, entry, activity, exit, spawn, 
           edge, walker, or, if, elif, else, for,  by, while,
           continue, break, disengage, report, anchor, has, can, true, false, from,
           context, info, details, try, strict, length, test, type, str,
           int, float, list, dict, bool, digraph, subgraph, test, by, 
           skip, assert, yield, class, def, Method, Function, pass,except,strip,split,
       },
   sensitive=false, 
   morecomment=[l]{//}, 
   morecomment=[l]{\#}, 
   morecomment=[s]{/}{/}, 
   morecomment=[s]{```}{```}, 
   moredelim=[is][\color{blue}]{\{}{\}}, 
   morestring=[b]", 
   morestring=[b]`,
   morestring=[s][]{\{}{\}},
   moredelim=[s][\color{orange}]{f"}{"}
}

\newcommand\pythonstyle{\lstset{
    language=Python,
    frame=single,                                    
    columns=fullflexible,
    breaklines=true,
    captionpos=t,
    basicstyle=\fontsize{6pt}{5pt}\upshape\ttfamily,      
    numbers=left,                    
    numbersep=8pt,                   
    numberstyle=\tiny\color{gray}, 
    keywordstyle=\bfseries\color{blue!55!black},
    stringstyle=\color{orange},
    xleftmargin=14pt, 
    xrightmargin=6pt,
    commentstyle=\color{gray}\ttfamily,
    backgroundcolor=\color{white},
    morekeywords={self},              
    emph={MyClass,__init__,return,type, str, int,
               int, float, list, dict, bool,tuple},          
    emphstyle=\bfseries\color{green!50!black},    
    frame=tb,                         
    showstringspaces=false
}

}

\lstnewenvironment{python}[1][]
{
\pythonstyle
\lstset{#1}
}
{}

\newcommand\pythoninline[1]{{\pythonstyle\lstinline!#1!}}

\definecolor{addition}{rgb}{0, 0.6, 0} 
\definecolor{deletion}{rgb}{0.8, 0, 0} 
\definecolor{unchanged}{rgb}{0, 0, 0}  

\lstnewenvironment{pythondiff}[1][]
{
    \pythonstyle
    \lstset{
        showspaces=false,                          
        showtabs=false,                            
        columns=fixed,                             
        keepspaces=true,                           
        morecomment=[f][\color{deletion}]{-},       
        morecomment=[f][\color{addition}]{+},       
        morecomment=[f][\color{unchanged}]{ },      
        moredelim=[il][\color{deletion}\sout]{\-},  
        moredelim=[il][\color{addition}]{\+},       
        literate=                                  
        {-}{{\color{deletion}-}}1                  
        {+}{{\color{addition}+}}1                  
        {^}{{\ }}1 
        #1 
    }
}
{}

\newcommand\jacstyle{
    \lstdefinelanguage{jac}{
       keywords={sem, import, node, ignore, take, activity, exit, spawn, with,
               edge, walker, and, or, if, elif, else, for, with, by, while,
               continue, break, disengage, report, anchor, has, can, true, false, from,
               context, info, details, try, strict, length, test, digraph, subgraph, test, by, in, to,
               skip, assert, yield, class, obj, Method, can, glob, from
           },
       sensitive=false, 
       morecomment=[l]{//}, 
       morecomment=[l]{\#}, 
       morecomment=[s]{/}{/}, 
       morestring=[b]", 
       morestring=[b]',
       morestring=[b]`,
       morestring=[s][]{\{}{\}}
    }
    
    \lstset{
        language=jac,
        frame=single,
        columns=fullflexible,
        breaklines=true,
        captionpos=t,
        basicstyle=\fontsize{6}{5}\upshape\ttfamily,
        numbers=left,
        numbersep=8pt,
        numberstyle=\tiny\color{gray},
        keywordstyle=\bfseries\color{blue!55!black},
        stringstyle=\color{orange},
        xleftmargin=5pt,
        xrightmargin=10pt,
        backgroundcolor=\color{white},
        morekeywords={self},
        emph={return,by,py,jac,entry,type, str, int,
               int, float, list, dict, bool},
        emphstyle=\bfseries\color{green!50!black},
        frame=tb,                         
        showstringspaces=false
    }

}

\newcommand\jacinline[1]{{\jacstyle\lstinline!#1!}}

\lstnewenvironment{jac}[1][]
    {
    \leavevmode
    \jacstyle
    \lstset{#1}
    }
{}

\lstdefinelanguage{output}{
    keywords={Output},
    basicstyle=\tiny\ttfamily, 
    breaklines=true,
    keywordstyle=\color{blue},
    stringstyle=\color{black},
    commentstyle=\color{gray},
    morecomment=[l]{//},
    columns=fullflexible,
    xleftmargin=15pt,
    xrightmargin=5pt,
    showstringspaces=false
}

\lstdefinelanguage{vision_output}{
    keywords={Output},
    basicstyle=\tiny\ttfamily, 
    breaklines=true,
    keywordstyle=\color{blue},
    stringstyle=\color{black},
    commentstyle=\color{gray},
    morecomment=[l]{//},
    columns=fullflexible,
    xleftmargin=5pt,
    xrightmargin=5pt,
    showstringspaces=false
}
\newcommand{\down}[1]{\textcolor{darkgreen}{\bm{$\downarrow$}} \fontsize{10pt}{10pt}\selectfont{#1}\fontsize{8pt}{8pt}\selectfont{$\times$}}

\begin{document}

\title{MTP: A Meaning-Typed Language Abstraction for AI-Integrated Programming}

\author{Jayanaka L. Dantanarayana}
\orcid{0009-0000-4320-8280}
\affiliation{%
  \institution{University of Michigan}
  \city{Ann Arbor}
  \country{USA}
}
\email{jayanaka@umich.edu}

\author{Yiping Kang}
\orcid{0000-0002-5964-3655}
\affiliation{%
  \institution{University of Michigan}
  \city{Ann Arbor}
  \country{USA}
}
\email{ypkang@umich.edu}

\author{Kugesan Sivasothynathan}
\orcid{0009-0004-4657-4947}
\affiliation{%
  \institution{Jaseci Labs}
  \city{Ann Arbor}
  \country{USA}
}
\email{kugesan.sivasothynathan@jaseci.org}

\author{Christopher Clarke}
\orcid{0000-0001-8741-3155}
\affiliation{%
  \institution{University of Michigan}
  \city{Ann Arbor}
  \country{USA}
}
\email{csclarke@umich.edu}

\author{Baichuan Li}
\orcid{0009-0008-4812-6303}
\affiliation{%
  \institution{University of Michigan}
  \city{Ann Arbor}
  \country{USA}
}
\email{patrli@umich.edu}

\author{Savini Kashmira}
\orcid{0009-0005-4911-7597}
\affiliation{%
  \institution{University of Michigan}
  \city{Ann Arbor}
  \country{USA}
}
\email{savinik@umich.edu}

\author{Krisztian Flautner}
\orcid{0009-0002-8347-1811}
\affiliation{%
  \institution{University of Michigan}
  \city{Ann Arbor}
  \country{USA}
}
\email{manowar@umich.edu}

\author{Lingjia Tang}
\orcid{0000-0002-5609-7775}
\affiliation{%
  \institution{University of Michigan}
  \city{Ann Arbor}
  \country{USA}
}
\email{lingjia@umich.edu}

\author{Jason Mars}
\orcid{0000-0002-7029-5292}
\affiliation{%
  \institution{University of Michigan}
  \city{Ann Arbor}
  \country{USA}
}
\email{profmars@umich.edu}

\renewcommand{\shortauthors}{J.L. Dantanarayana, Y. Kang, K. Sivasothynathan, C. Clarke, B. Li, S. Kashmira, K. Flautner, L. Tang, and J. Mars}
\begin{abstract}

Software development is shifting from traditional programming to \textit{AI-integrated} applications that leverage generative AI and large language models (LLMs) during runtime. However, integrating LLMs remains complex, requiring developers to manually craft prompts and process outputs. Existing tools attempt to assist with prompt engineering, but often introduce additional complexity.

This paper presents \textbf{Meaning-Typed Programming (MTP)}, a novel paradigm that abstracts LLM integration through intuitive language-level constructs. By leveraging the inherent semantic richness of code, MTP automates prompt generation and response handling without additional developer effort. We introduce the \textbf{(1) by} operator for seamless LLM invocation, \textbf{(2) MT-IR}, a meaning-based intermediate representation for semantic extraction, and \textbf{(3) MT-Runtime}, an automated system for managing LLM interactions. We implement MTP in \textbf{Jac}, a programming language that supersets Python, and find that MTP significantly reduces coding complexity while maintaining accuracy and efficiency.  MTP significantly reduces development complexity, lines of code modifications needed,
and costs while improving run-time performance and maintaining or exceeding the accuracy
of existing approaches. Our user study shows that developers using MTP completed tasks 3.2× faster with 45\% fewer lines of code compared to existing frameworks. Moreover, \systemname{} demonstrates resilience even when up to 50\% of naming conventions are degraded, demonstrating robustness to suboptimal code. \systemname{} is developed as part of the Jaseci open-source project, and is available under the module \href{https://github.com/jaseci-labs/jaseci/tree/main/jac-byllm}{\textbf{byLLM}}.

\end{abstract}






\maketitle

\section{Introduction}

With the advent of generative AI and Large Language Models (LLMs), the way we develop software is rapidly evolving. While much work has focused on using LLMs in \textit{code generation}~\cite{wang2023review, levin2025pythoness}, an emerging trend is the integration of LLMs as components that provide essential functionality~\cite{sharma2025promptpex, Beurer2023LMQL, li2023scallop, weber2024large}. These \emph{AI-Integrated applications} leverage generative AI at runtime to perform critical program functionality, merging conventional programming with AI-driven capabilities. This shift fundamentally transforms software development, redefining how applications are designed and built.

\begin{figure}[t]
    \centering
    \includegraphics[width=\linewidth]{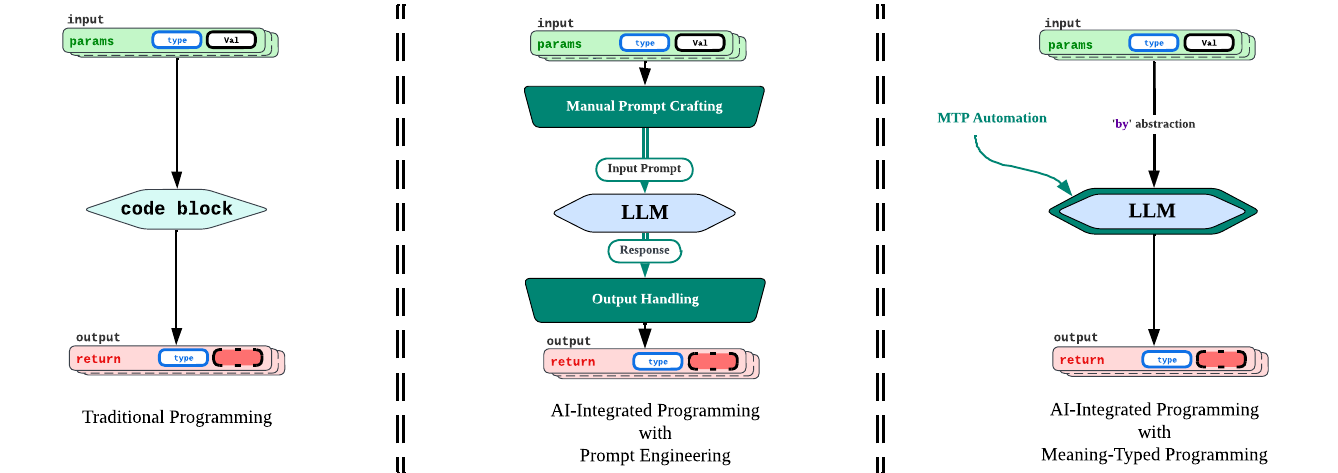}
    \caption{High level view of what traditional program functionality consists of (Left) vs. AI-Integrated programming using prompt engineering (Middle) and Meaning-Typed Programming(\systemname{}) paradigm (Right).}
    \label{fig:meaning-type-intro}
    \Description{}
\end{figure}

Building AI-Integrated applications remains a complex challenge due to the fundamental differences between operating assumptions. In conventional programming, code describes precise operations performed on explicitly defined variables such as objects (Figure~\ref{fig:meaning-type-intro} left). In contrast, LLMs process natural language text as input and produce text-based outputs. To integrate LLMs into programs, developers must manually construct textual input, a process known as \emph{prompt engineering}.
Prompt engineering is the primary method of realizing AI-Integrated programming and can be tedious, complex, and imprecise. Developers must craft descriptive instructions, manage textual context, and handle the inherent variability in LLM output. This process includes a significant amount of trial-and-error, making it time-consuming and difficult to maintain.


Several efforts have been made to address these challenges by developing frameworks and tools to facilitate prompt engineering. Frameworks such as LangChain~\cite{langchain}, Language Model Query Language (LMQL)~\cite{Beurer2023LMQL}, DSPy~\cite{khattab2023dspy}, and SGLang~\cite{zheng2023SGLang} have been introduced to assist developers in generating and managing prompts. However, while these frameworks provide more sophisticated tooling to help bridge the gap between conventional programming and the incorporation of AI models, in practice, they introduce additional layers of complexity for developers:

\begin{enumerate} [nosep]
    \item \emph{Prompt design complexity:} Developers remain responsible for the manual construction of prompts, including determining the appropriate prompt language and selecting relevant information to include. 
    \item \emph{Steep learning curve:} These systems require developers to familiarize themselves with new query languages, frameworks, or specialized syntax, complicating the integration process.
    \item \emph{Input/output conversion complexity:} Parsing LLM output and converting them back to objects remains a challenge, especially given the variability in output across different LLMs.
\end{enumerate}

In this paper, we introduce \emph{meaning-typed programming} (MTP), a novel approach to simplify the creation of AI-Integrated applications by (i) introducing new high level language abstractions that hide the complexities of \emph{AI-Integration}, (ii) a novel technique for automating prompt generation using inherent code semantics, and (iii) a runtime system that automates LLM inference and output handling.

 The \textbf{key insights} of this work are: \emph{(1) In practice, programs are written to be readable by various developers to comprehend the intentions and semantics conveyed by the code. This semantic richness can be leveraged to automatically translate intent embodied in the code to prompts for the LLM. (2) As state-of-the-art LLMs continue to advance, AI becomes increasingly feasible to infer code intentions without requiring explicit descriptions or prompt construction by developers. This trend is not only supported by our empirical findings, but we surprisingly find less can perhaps be more in prompt engineering (see \S~\ref{sec:Eval}).}

There are four primary \textbf{challenges} that must be overcome in the design and implementation of MTP. Firstly, the introduced language abstraction should be simple, intuitive, and flexible. Secondly, developing an automated approach that provides LLMs with sufficient semantic information for accurate task performance is challenging. Thirdly, since AI-Integrated applications rely on dynamic values, prompts must be generated at runtime, requiring a lightweight system to handle prompt generation and LLM inference efficiently. Lastly, because LLMs generate non-deterministic outputs, the outputs must be carefully processed to ensure correct type conversion. To address these challenges, we introduce three key components to enable the seamless creation of AI-Integrated programs. These include 

\begin{enumerate}

    \item \textbf{The `by' operator}, that enables the seamless integration of LLM functionality into code. It acts as a bridge between traditional and LLM operations, allowing developers to replace function bodies with a simple \byop{} \textbf{model\_name} clause.
\begin{footnotesize}
\begin{python}
def calc_age(birth: date, today: date) -> int by gpt4()
\end{python}
\end{footnotesize}
As shown later in our evaluation, meticulously extracted relevant semantic information is sufficient for the LLM to produce the accurate result, significantly simplifying LLM integration into existing codebases. 
    \item \textbf{MT-IR}, a meaning-based intermediate representation at the compiler level that collects and organizes semantically rich information, preserving the meaning behind variable names, function signatures, and types, during runtime. 
    \item \textbf{MT-Runtime}, an automated runtime engine within the language's virtual machine, triggered at \texttt{by} call sites. It binds runtime values to MT-IR dynamically, manages LLM interactions, generates context-aware prompts, and handles responses and errors. Through these three components, \systemname{} allows developers to focus on application logic while automatically managing LLM integration, ensuring seamless and contextualized operations.
\end{enumerate}
We implement meaning-typed programming (MTP) in a production-grade Python superset language called Jac~\cite{mars2023jaseci, jaseci-repo, mars2025objectspatialprogramming}. The \byop{} operator was introduced as a language primitive in Jac, and the MT-IR construction is implemented as a compilation pass in Jac's ahead-of-time compiler. The MT-Runtime system is implemented as a language plugin which can be separately installed as a PyPi package. This end-to-end framework has been used to develop real-world applications and benchmarks utilized in our extensive evaluation.

Specifically, we make the following contributions:

\begin{itemize} [nosep]
    \item We introduce \textbf{meaning-typed programming}, a novel paradigm that leverages the inherent semantic richness of code to automate LLM integration into applications without requiring explicit prompt engineering.
    

\item We introduce the \textbf{\byop{} operator}, a simple yet powerful language abstraction that provides both intuitive and flexible integration of LLM functionality into code. It enables developers to seamlessly incorporate language models for multiple code constructs including functions, methods, and object initialization while maintaining familiar programming patterns.

    \item We develop supporting compiler and runtime infrastructure: \textbf{MT-IR}, a semantically rich intermediate representation that extracts and organizes code meaning, and \textbf{MT-Runtime}, an automated execution engine that dynamically converts semantic information into prompts, manages LLM interactions, and handles type-safe response processing.

    \item We \textbf{implement MTP in Jac}, a production-grade Python superset language, conduct extensive evaluation and show that MTP significantly reduces development complexity, lines of code, and costs while improving run-time performance and maintaining or exceeding the accuracy of existing approaches.

    \item We present multiple \textbf{case studies} demonstrating the application of MTP in complex problem domains, analyzing its impact on code structure, development workflow, and complexity reduction compared to traditional LLM integration methods.

\end{itemize}

\begin{wrapfigure}{r}{0.4\textwidth} 
    \centering
     \vspace{-0.2cm}
    \input{code/data_structures}
    \caption{A typical implementation for a video game level development usecase. Code shows how the data-structure definitions for the video game application.}
    \label{fig:video_game_datastructures}
    \input{code/non_NI_game}
        \caption{Traditional implementation of game levels for the non AI-Integrated usecase.}
    \Description{}
    \label{fig:video_game_trad}
    \vspace{-0.5cm}
\end{wrapfigure}

We perform extensive evaluation to investigate the implications of \systemname{} on code complexity, AI-Integration accuracy, inference token usage, runtime performance, and cost compared to prior work. We also assess how sensitive \systemname{} is to poor coding practices, as our system is built on leveraging semantically rich code. Our evaluations show that MTP reduces lines of code by factors of 2.3-7.5× compared to LMQL and 1.3-10.7× compared to DSPy, demonstrating significant code complexity reduction. In terms of accuracy, MTP performs on par with or better than DSPy, while surpassing LMQL on average by 8\% across all benchmarks when evaluated with gpt-4o model. For accuracy on math problems from the GSM8k~\cite{cobbe2021gsm8k} dataset, \systemname{} outperformed other frameworks with newer OpenAI and Llama models, achieving accuracies approaching 90\%. Regarding inference costs, \systemname{} demonstrated savings over DSPy on most benchmarks, with reductions up to 4.5×. Moreover, \systemname{} enables program speedups of up to 4.75× over DSPy, indicating minimal runtime overhead. Our sensitivity studies confirm that MTP maintains effectiveness even when 50\% of naming conventions are suboptimal. These results establish MTP as a practical, efficient solution for streamlining AI-Integrated application development.

\section{Motivation Example}
\label{sec:motivating_example}

This section presents a motivating example to illustrate the concept of \textit{AI-Integrated} programming and to highlight the complexities involved in the current implementation technique, i.e., prompt engineering. The example application used here is a video game implementation, where dynamically generating game levels is achieved through AI-Integration with an LLM.

The section is organized as follows. We first introduce the example application and the code sketch of traditional implementation without an LLM (Figure \ref{fig:video_game_trad}). We then demonstrate AI-Integrated implementation using the industry-standard approach of prompt engineering (Figure \ref{fig:video_game_PE}), highlighting the significant overhead it imposes on developers. Lastly, we present implementation using our proposed MTP approach (Figure \ref{fig:video_game_mtp}), demonstrating MTP's capabilities in removing complexities associated with prompt engineering.
\vspace{-0.6em}

\begin{wrapfigure}{l}{0.5\textwidth}
    \vspace{-0.4cm} 
    \centering
    \input{code/Game_with_PE}
    \caption{Implementation of the video game level generation using prompt-engineering}
    \Description{}
    \label{fig:video_game_PE}
    \vspace{0.5cm}
\end{wrapfigure}

\subsection{Traditional Programming}
The video game implementation used here is a real program developed based on the pygame~\cite{mcgugan2007beginning} tutorials. In this video game, the player must complete the current game level to proceed to the next. An example of the video game level progression is shown in \S~\ref{sec:Eval} (Figure~\ref{fig:rpg_levels}), where each level includes a set of configurations (maps, enemies, difficulties, etc). Figure~\ref{fig:video_game_datastructures} presents the basic data structures used in the video game implementation, including the \texttt{Level} object (lines 14–21), which encapsulates the configurations and a nested \texttt{Map} object (lines 9–12).

\textbf{\texttt{get\_next\_level} } The level progression logic is achieved through \texttt{get\_next\_level} function, which uses the previous level as input and returns the next level. Figure~\ref{fig:video_game_trad} shows the code sketch, including a placeholder (lines 2–6) for the function body. In traditional programming, developers must implement complex algorithms and logic to generate the next level based on the current one, often leveraging procedural generation techniques, which can be both challenging and time-consuming. However, by integrating an LLM, a level can be generated automatically without developers' explicit coding.

\vspace{-0.2cm}
\subsection{State-of-the-Art AI-Integration Approach}
\label{sec:NIPE}

\textbf{C\textit{ode Synthesis vs. AI-Integration.}} Before presenting the current approach to AI-Integration, it is important to highlight the difference between AI-Integration and the use of LLMs for code synthesis~\cite{wang2023review}. LLM-enabled code synthesis uses LLMs to generate code that is executed at runtime. In contrast, \emph{AI-Integrated} code~\cite{weber2024large} leverages an LLM to dynamically perform tasks within a program. During runtime, the LLM inference generates output values based on given inputs to the LLM. In this example, \emph{AI-Integrated} approach uses the input (previous level object) to query an LLM at runtime. The output of the LLM inference  will be used as the next level configuration.

Current approaches to AI-Integration primarily rely on prompt engineering as the industry standard, which involves crafting input prompts to guide LLMs in generating desired responses. It involves structuring inputs, optimizing the wording and context of the prompt to maximize the accuracy and relevance of the model's output. Figure \ref{fig:video_game_PE} presents the prompt engineering-based implementation of {get\_next\_level} functionality, where the input of the previous level object is described in a manually-craft prompt to an LLM, and the output of the LLM (in a text format) is used to construct the next level. The LLM innovation is in lines 37-42, which eliminates the need to manually implement the level generation algorithms. However,  this example illustrates the significant complexity surrounding the LLM invocation/integration. 

 \vspace{-0.28cm}
\paragraph{\textbf{Manual Prompt Crafting}}
 The example includes a manually crafted prompt template (lines 7-35). As shown here, to effectively improve LLM accuracy, the prompt consists of several key components: a problem description(line 8), an input object (including a description of the input and the \texttt{prev\_level} object in JSON format), and a set of rules (lines 14-18) to guide the LLM's generation. In addition, developers need to explicitly specify the output format(lines 20-34), which in this case follows a JSON schema for the \texttt{Level} object. Since the \texttt{Level} object contains deeply nested objects, structuring the output format is tedious and labor-intensive. This prompt engineering process is time-consuming and often requires an iterative trial-and-error process. 

 \vspace{-0.28cm}
\paragraph{\textbf{Output Type Conversion}}
In addition to prompt crafting, developers must also convert the LLM's output into a return-typed object compatible with traditional programming. In this example, LLM's output follows the JSON object format, as specified in the prompt, which then must be converted into a \texttt{Level} object that is understandable by the rest of the application(lines 47-70). This includes mapping the JSON output to the corresponding attributes of the \texttt{Level} class (lines 62-70). To handle deeply nested objects, three additional helper functions are defined on lines 47, 50, and 56. This added complexity makes AI-Integration with prompt engineering more cumbersome and less intuitive.

\vspace{-0.2cm}
\paragraph{\textbf{Error Handling and Retries}}
Due to the non-deterministic nature of LLMs, the generated output may not always be consistent with the expected format, leading to program runtime errors that developers must anticipate and handle. In addition to error-handling, a retry mechanism may be needed to reattempt the LLM invocation process until a correctly formatted output is produced. Although the example code snippet does not include error handling and retries due to space constraints, incorporating such mechanisms adds additional complexity and requires development effort to ensure robustness and reliability.

In conclusion, this example demonstrates that while AI-Integration can leverage LLMs to improve traditional programming, the complexities associated with prompt engineering may introduce significant cognitive and implementation burden. Several frameworks and tools have been developed to simplify the interaction with LLMs. While these systems offer
valuable capabilities for prompt engineering and LLM integration, they usually incur additional complexity. LMQL~\cite{Beurer2023LMQL} requires developers to learn a new query language, while DSPy~\cite{khattab2023dspy} requires the developers to annotate all essential information in code, replacing prompt-engineering complexities with annotation overhead. The comparison to prior work is presented later in \S~\ref{sec:Eval}. 

\vspace{-0.15cm}
\subsection{MTP Vision for AI-Integrated Programming}

To address the challenges in the current AI-Integration approach, our research aims to \textit{design a language abstraction and system to streamline the process of integrating LLMs into traditional programming by abstracting and automating prompt generation, output type conversions, and error handling.}

\begin{wrapfigure}{r}{0.50\textwidth} 
    \centering
    \input{code/MTP_game}
        \caption{Implementation of AI-Integration as envisioned in this paper, with Meaning-Typed programming.}
    \Description{}
    \label{fig:video_game_mtp}
    \includegraphics[width=1\linewidth]{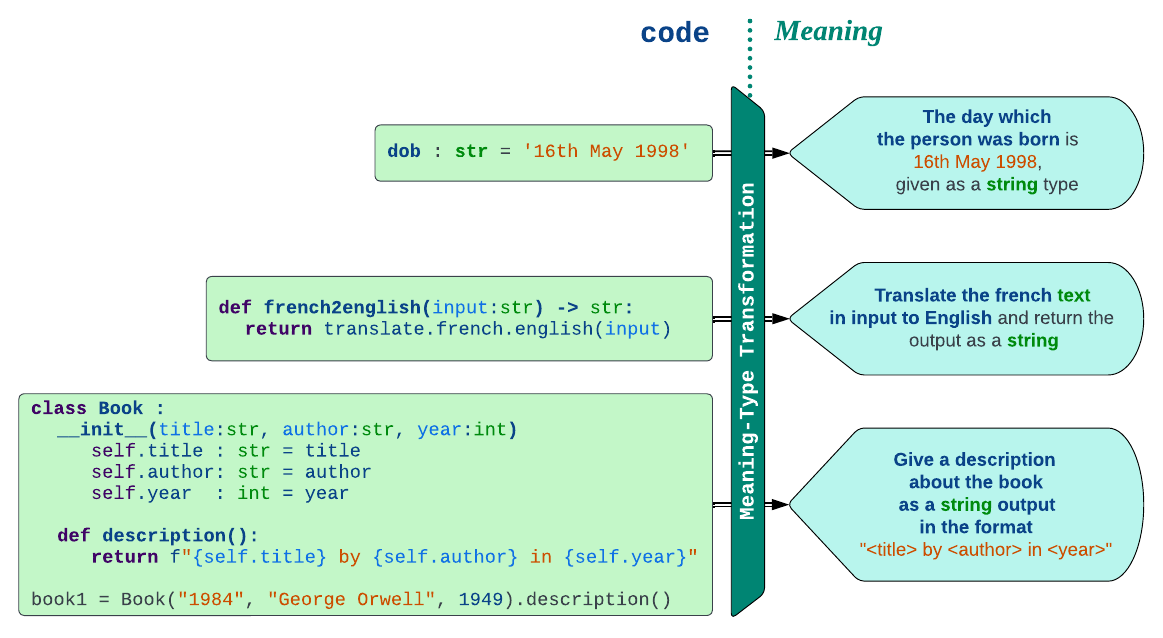}
    \caption{Examples of~\textit{meaning} embedded in code semantics.}
    \Description{}
    \label{fig:meaning}
\end{wrapfigure}

Figure \ref{fig:video_game_mtp} illustrates our envisioned approach, where adding only a few lines of code enables LLM integration, eliminating the need for prompt engineering and its associated complexities. Moreover, this approach is significantly more intuitive than prior works. The code snippet provides a preview of what we introduce in this paper as \textit{Meaning-Typed Programming}. Line 5 of the code snippet shows how AI-Integration is achieved. \systemname{} hides prompt-engineering using a language abstraction, automating the prompt generation, LLM inference and output handling. We will describe details on our language abstraction in \S~\ref{sub: by}.

When hiding prompt engineering, the information required for accurate LLM inference should be available to generate prompts. As shown in Figure~\ref{fig:video_game_mtp}, the intent of the \texttt{get\_next\_level()} function on line 5 is evident from the code’s semantics. This observation leads to the key insight of our paper: \textit{programs are written to be readable by various developers to comprehend the intentions and semantics conveyed by the code. This semantic richness can be leveraged to automatically translate intent embodied in the code to prompts for the LLM.}

Figure~\ref{fig:meaning} presents three additional illustrative examples showing how meaning can be derived from various code elements:
\begin{enumerate}
\item The meaning of the variable \texttt{dob} can be inferred from its name, type, and value. Here \texttt{dob} is a commonly used abbreviation for date of birth which still retains adequate semantic meaning.
\item The meaning of the \texttt{french2english()} function can be inferred from its name, input parameter, and return type. These semantics indicate the function takes French text as input and returns an English translation of that text. 
\item The semantics of the \texttt{description()} method in the \texttt{book} class can be understood by analyzing the class, variable names, and an example function call.
\end{enumerate}

With the help of a compiler and run-time techniques, LLMs can infer the intended meaning from a program to perform AI-Integration tasks as shown in our evaluations, later in the \S \ref{sec:Eval}. %

\section{Problem, Objectives and Challenges}
\label{sec:problem}

Our goal is to leverage semantically rich, human-readable code expressions to automate prompt-engineering. Instead of requiring developers to manually craft detailed prompts or annotations, we harness the inherent structure and semantics of the code itself to generate meaningful LLM inputs.

To achieve this goal, we define \textbf{\textit{three key objectives}}:

\begin{enumerate}[label=\textbf{O$_\arabic*$}]
\item \label{O1} - Design simple, intuitive and flexible language abstractions for developers while hiding prompt engineering and other complexities.
\item \label{O2} - Design techniques to extract relevant and sufficient semantic information from code for automatic prompt synthesis.
\item \label{O3} - Design a system that combines semantic information and dynamic information to manage prompt synthesis and output interpretation.
\end{enumerate}

When designing a solution that achieves these objectives, \textbf{\textit{four unique challenges arise}}:
\begin{enumerate}[label=\textbf{C$_\arabic*$}]
\item \label{C1} - How do we design a language abstraction that supports multiple AI-Integration methods while remaining easy to learn and use, minimizing the developer learning curve? 
\item \label{C2} - What semantic information must be extracted to generate an LLM prompt that ensures program accuracy without incurring cost and runtime overhead? 
\item \label{C3} - What type of runtime system design is appropriate to combine static semantic information with dynamic information to manage LLM input?
\item \label{C4} - How do we design a system that handle output type conversion and error handling without increasing token cost and runtime cost?
\end{enumerate}

\ref{C1} involves designing a simple and intuitive language abstraction, minimizing the learning curve and cognitive load. At the same time, it must be flexible enough to seamlessly integrate across different code locations. Designing a system that is both intuitive and flexible is challenging, as the abstraction must be simple enough to reduce the learning curve while also being usable orthogonally at multiple AI-Integration points.

\ref{C2} involves extracting only the relevant semantics for AI-Integration points. A naive approach would be to include the entire source code in the prompt, but this would lead to high token costs, or even context window overflow. To optimize token usage, we must selectively extract only the relevant semantics. But this is particularly challenging because these semantics are often scattered across different code locations, requiring sophisticated codebase-wide analysis to identify and extract only what is necessary for accurate LLM prompting.

\ref{C3} involves combining extracted code semantics with dynamic values to create effective prompts. This requires accessing variable data at runtime, which can vary in complexity depending on where the integration happens. For functions, it is straightforward since values are passed as parameters. However, for object methods, it becomes more complex because the system must also access relevant object attributes at runtime. To handle this, a sophisticated runtime system is needed to capture the execution context and combine dynamic values into the prompts.

\ref{C4} involves correctly interpreting LLM-generated responses and converting them to appropriate types with robust error handling. This is particularly difficult because LLMs generate outputs non-deterministically and often deviate from expected formats. As demonstrated in the prompt-engineering example in \S \ref{sec:NIPE}, extracting correct results requires careful parsing, but inconsistencies in output structure make this process complex. Implementing a retry mechanism to correct such inconsistencies requires additional LLM inferences, increasing token usage and cost. Therefore, the system design must balance effective error handling with minimal impact on token consumption and runtime cost.

\vspace{-0.15cm}
\section{Meaning Typed Programming}
\label{sec:methodology}

\begin{figure}[t]
        \centering
        \includegraphics[width=0.9\linewidth]{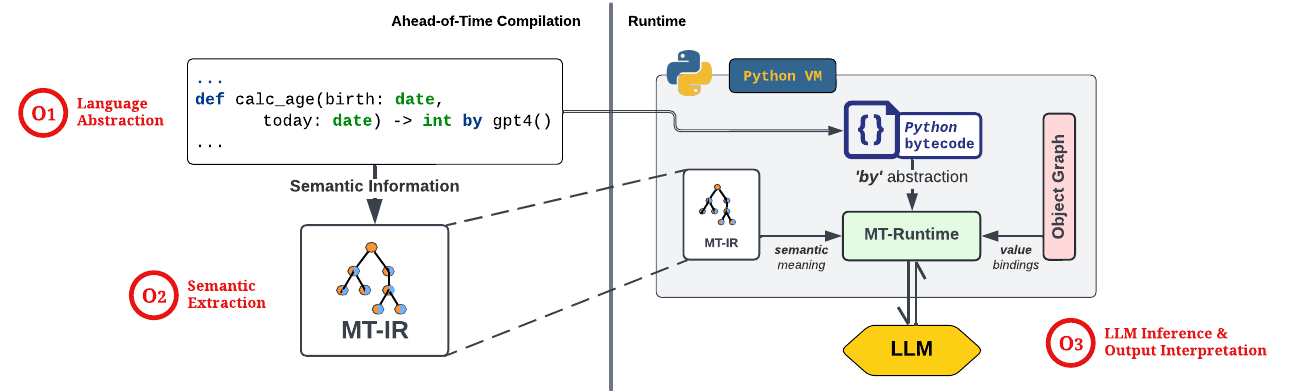}
\caption{Meaning-Typed Programming system overview.}
\Description{}
\label{fig:MTP}
\end{figure}

To address the problem of automating and streamlining AI-Integrated application development discussed in \S \ref{sec:problem}, we introduce \textit{Meaning Typed Programming (MTP)}. 
The MTP paradigm (Figure~\ref{fig:MTP}) comprises three key components that achieves each objective defined in \S \ref{sec:problem}, enabling seamless AI-Integration into programming languages:

\begin{enumerate} [nosep]
    \item The \byop{} operator as the language level abstraction. (To achieve \ref{O1})
    \item Meaning-Typed Intermediate Representation (MT-IR), generated during compilation. (To achieve \ref{O2})
    \item MT-Runtime, an automated runtime engine that manages the LLM integration. (To achieve \ref{O3})
\end{enumerate}
This section presents an overview of MTP as well as an in depth view into each component introduced in this paper with respect to the challenges mentioned in \S \ref{sec:problem}.

\vspace{-0.15cm}
\subsection{MTP Overview}

\textbf{The \byop{} operator} acts as a bridge between traditional code and LLM operations, allowing developers to seamlessly integrate LLM functionality into their code by simply adding the \byop{} keyword. As shown in Figure~\ref{fig:MTP}, developers can replace a function’s implementation, such as \texttt{calc\_age()}, with a straightforward \byop{} <\texttt{model\_name}> clause. At runtime, the \byop{} operator invokes MT-Runtime to execute the desired functionality using the specified LLM, achieving \ref{O1}. In this example, the system leverages GPT-4 to calculate age based on a given birthday and the current date. \S~\ref{sub: by} provides further details on the \byop{} operator.

\textbf{MT-IR} is an intermediate representation designed to capture and preserve the meaning of the program to be used at runtime upon invocation of the \byop{} keyword. During compilation, MT-IR is generated by analyzing and organizing semantically rich information in the code (see Figure~\ref{fig:MTP} (left)). It preserves key elements such as variable names, function signatures, and object schema relevant to a given \byop{} call-site to provide access to the code's original meaning by the LLM at run-time (see Figure~\ref{fig:MTP} (right)) achieving \ref{O2}. 
(see \S~\ref{sec:MTIR} for more details). 
 
 \textbf{MT-Runtime} manages the complexities of LLM interaction, through prompt synthesis using MT-IR. It enhances execution by intelligent response parsing, and dynamic error handling, achieving \ref{O3}. It operates within the language's virtual machine (e.g., Python's VM) and is specifically triggered at \byop{} call sites. MT-Runtime combines both static information from MT-IR and dynamic values from the language's object graph. By seamlessly integrating with the language's execution environment, MT-Runtime allows developers to focus on their application logic while automating LLM interaction (see \S~\ref{sec:MT-Runtime} for more details). 

\subsection{Meaning-Type Language Construct: `\byop{}' Operator}
\label{sub: by}

Hiding prompt engineering within \textit{a language abstraction that is simple, intuitive and flexible is challenging} as discussed in \S \ref{sec:problem} under \ref{C1}. To address this we introduce the \byop{} operator as the core language abstraction of MTP. The base syntax of the \byop{} operator is,

\begin{center}
  \textit{<code construct>} \byop{} \textbf{\textit{llm\_ref}}( \textit{model\_hyperparameters} )
\end{center}

Here, the workload defined by the code construct on the left-hand side of the \byop{} operator is AI-Integrated using the LLM specified on the right. This syntax is simple and intuitive, making it easy to learn. Additionally, this syntax provides flexibility by enabling developers to integrate LLMs at multiple points within standard code constructs. Furthermore, additional hyperparameters can be provided as arguments to the LLM object for further customization.

In this paper, we introduce three code construct locations to use the \byop{} operator, allowing developers to use MTP flexibly while maintaining simplicity and minimal overhead. These locations include \textit{function definitions}, \textit{object initialization} and \textit{member method definitions}.

To formally express the semantic representation of \byop{} operator usage across various integration points and their respective call sites, we introduce a formal semantics that captures the operational behavior of \byop{} in a principled way. We define the following notation:

\begin{itemize}[nosep]
    \item $\mathcal{M}$: the set of available language models.
    
    \item $eval( \cdot )$: the semantic evaluation function.
  
\end{itemize}

These definitions serve as the foundation for describing how \byop{} transforms traditional code constructs into AI-Integrated operations.

\subsubsection{\byop{} at Function Definitions}
Consider a standalone function $f$ with parameters $p_1, p_2, \ldots, p_n$ of types $T_1, T_2, \ldots, T_n$ respectively, and having output type $T_r$. Using the \byop{} operator, this function can be integrated with an LLM $m \in \mathcal{M}$ with hyperparameters $\theta$ as follows:

\begin{equation}
\label{eq:function-def-rule}
\text{def } f(p_1:T_1, \ldots, p_n:T_n) \rightarrow T_r \text{ \byop{} } m(\theta)
\end{equation}

At runtime, invoking this function with actual arguments  $v_1, \ldots, v_n$ to return $v_r$ of type $T_r$, is semantically equivalent to:
\begin{equation}
\label{eq:function-call-rule}
v_r = eval( f(v_1, \ldots, v_n) ) = \text{invoke-model}(m, \theta, f, [v_1, \ldots, v_n], [(T_1, \ldots, T_n), T_r])
\end{equation}

Here, the model ($m$) receives the function signature $f$, the actual arguments $(v_1, \ldots, v_n)$ with their corresponding types $(T_1, \ldots, T_n)$, and the expected output type $T_r$. By leveraging this rich semantic context, the model can reason about the intended behavior of the function and generate an appropriate output conforming to $T_r$. An illustrative example is provided in Figure~\ref{code: func-call}, where the function \texttt{calculate\_age} is defined to return an integer output by computing a person's age given their date of birth (\texttt{dob}) and the current year (\texttt{cur\_year}). 

\subsubsection{\byop{} at Object Initialization}
The \byop{} operator can also be used during object initialization to let an LLM fill in missing attribute values. Consider a class $C$ with attributes $a_1, \ldots, a_n$, each having types $T_1, \ldots, T_n$. If only the first $k$ attributes are provided during initialization ($k<n$), the remaining attributes can be completed by $m \in \mathcal{M}$  with $\theta$ hyperparameters, as shown below:

\begin{equation}
C(a_1, \ldots, a_k) \text{ \byop{} } m(\theta)
\end{equation}

At runtime, when such an object is created, the system calls the model with the class name $C$, the values of the attributes that were provided $(v_1, \ldots, v_k)$, their types $(T_1, \ldots, T_k)$ and the types of the remaining attributes $(T_{k+1}, \ldots, T_n)$. The model then uses this information to infer and generate appropriate values for the missing attributes to fully instantiate an object $obj$ of type $C$.
\begin{equation}
\label{eq:object-init-rule}
\resizebox{0.95\columnwidth}{!}{$
obj = eval( C(v_1, \ldots, v_k) \text{ \byop{} } m(\theta)) 
    = \text{invoke-model}(m, \theta, C, [v_1, \ldots, v_k], [(T_1, \ldots, T_k), (T_{k+1}, \ldots, T_n)])
$}
\end{equation}

As per the example shown in Figure \ref{code: object-init}, the \byop{} call uses the partially initialized object (name being "Einstein") to predict and fill in the remaining attributes (dob). \systemname{} generates a model prompt using the class name, attribute names, types, and the provided values, and ensures that the model’s output is converted into the correct types using MT-Runtime. This allows for clean and semantically guided object completion even when only partial input is given.

\subsubsection{\byop{} at Member Method Definition}
The \byop{} operator can also be used to define methods within a class. Consider a method $mth$ within a class $C$, which takes parameters $p_1, \ldots, p_n$ with types $T_1, \ldots, T_n$, and returns a value of type $T_r$. This method can be implemented using $m \in \mathcal{M}$, configured with hyperparameters $\theta$, as follows.

\begin{equation}
\label{eq:method-def-rule}
\textit{class } C \text{: def } mth(p_1:T_1, \ldots, p_n:T_n) \rightarrow T_r \text{ \byop{} } m(\theta)
\end{equation}

At runtime, when this method is called on an object instance $obj$ with input values $v_1, \ldots, v_n$, evaluation to return $v_r$ of type $T_r$, proceeds as:

\begin{equation}
\label{eq:method-call-rule}
\resizebox{0.95\columnwidth}{!}{$
v_r = eval( obj.mth(v_1, \ldots, v_n) ) = \text{invoke-model}(m, \theta, C.mth, obj, [v_1, \ldots, v_n], [C, (T_1, \ldots, T_n), T_r])
$}
\end{equation}

\begin{wrapfigure}{r}{0.44\textwidth}
\vspace{-0.3cm}
    \centering
    \begin{subfigure}[b]{0.44\textwidth}
        \begin{python}
def calculate_age (cur_year: int, dob: str) -> int by llm

age = calculate_age(2024,"March 14, 1879")
        \end{python}
        \caption{Function Call}
        \label{code: func-call}
    \end{subfigure}
    \hfill
    \begin{subfigure}[b]{0.44\textwidth}
        \begin{python}
class Person :
    name : str
    dob : str

einstein = Person("Einstein") by llm
        \end{python}
        \caption{Object Initialization}
        \label{code: object-init}
    \end{subfigure}
    \hfill
    \begin{subfigure}[b]{0.44\textwidth}
        \begin{python}
class Person:
    name : str
    dob : str
    
    def calculate_age (cur_year: int) -> int by llm(temperature=0.7)

einstein = Person("Einstein", "March 14, 1879")
einstein.calculate_age(2024)
        \end{python}
        \caption{Member Method Call}
        \label{code: method-call}
    \end{subfigure}
    \caption{Meaning-Typed Programming supports three AI-Integration points.}
    \Description{}
    \label{fig:code_constructs}
\end{wrapfigure}

Unlike standalone function calls, class methods have access to the instance’s internal state, which provides important context for the model. During evaluation, the model is given the method signature $C.mth$, the object instance $obj$, the input values and their types, and the expected return type. The object includes all class attributes along with their current values and type information, allowing the model to use this internal context in its reasoning. For example, in Figure~\ref{code: method-call}, the \texttt{calculate\_age} method of the \texttt{Person} class derives the \texttt{dob} attribute directly from the \texttt{einstein} object rather than requiring it as an argument.

The \byop{} operator provides specific behavioral guarantees through MT-IR (\S~\ref{sec:MTIR}) and MT-Runtime (\S~\ref{sec:MT-Runtime}):

\begin{enumerate}
    \item \textbf{Type Safety}: Output type conformance to $T_r$ is enforced for all LLM-generated results. When model output cannot be converted to the expected type $T_r$, the system raises a type error.
    \item \textbf{Temporal Behavior}: The \byop{} operator exhibits distinct execution patterns based on integration point. For functions and methods, definition time tags the construct for MT-IR processing, while call time triggers MT-Runtime execution. For object initialization, the \byop{} operator executes immediately during instantiation.
\end{enumerate}

\begin{figure}[b]
    \centering
    \begin{minipage}{0.4\textwidth}
        \centering
        \begin{subfigure}{\linewidth}
            \includegraphics[width=\linewidth]{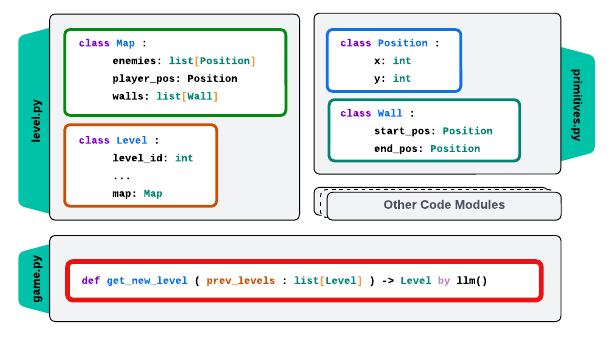}
            \caption{}
            \label{fig:game_code}
        \end{subfigure}
        
        \vspace{5pt} 

        \begin{subfigure}{\linewidth}
            \includegraphics[width=\linewidth]{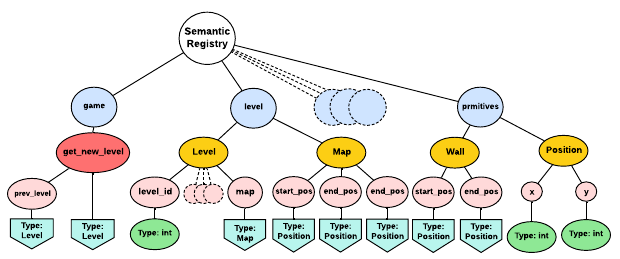}
            \caption{}
            \label{fig:semantic_registry}
        \end{subfigure}
    \end{minipage}
    \begin{minipage}{0.2\textwidth}
        \centering
        \begin{subfigure}{\linewidth}
            \includegraphics[width=\linewidth]{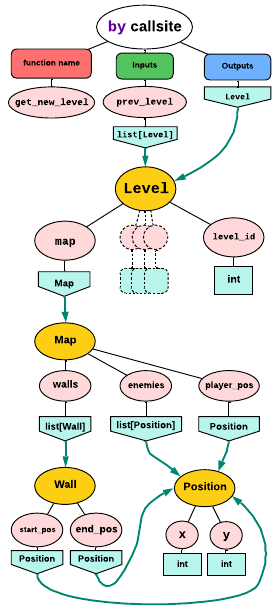}
            \caption{}
            \label{fig:MTIR_for_game}
        \end{subfigure}
    \end{minipage}
    \begin{minipage}{0.38\textwidth}
        \centering
        \begin{subfigure}{\linewidth}
            \includegraphics[width=\linewidth]{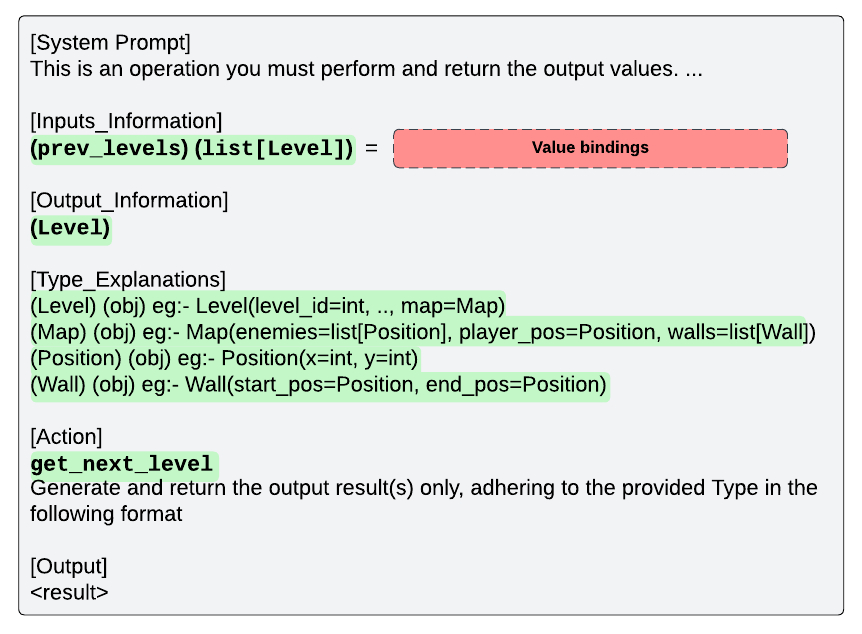}
            \caption{}
            \label{fig:genarated prompt}
        \end{subfigure}
    \end{minipage}

    \caption{Representation of how the MTIR is generated and used for the video game example in \S \ref{sec:motivating_example}. (a) Code layout in modules. (b) Corresponding semantic registry containing all semantic information in the codebase. (c) MTIR relevant for the \texttt{get\_next\_level} \byop{} call-site. (d) Generated prompt using the MTIR (semantic information extracted using MTIR are highlighted in green). }
    \label{fig:MTIR genaration and prompt integration}
\end{figure}

\subsection{Meaning-Typed Compilation and IR (MT-IR)}
\label{sec:MTIR}

When a \byop{} call site is encountered, we perform static analysis to extract relevant code semantics, which are used later at runtime to automatically generate prompts for the LLM inference. These relevant semantics, such as object types and function/argument signatures, are available at the source code level and might be lost at the bytecode level. Therefore, they must be captured ahead of time to enable accurate inference during the runtime. As discussed in \S \ref{sec:problem} \ref{C2}, this task is challenging because the relevant semantics are often distributed across multiple source files. Figure \ref{fig:game_code} illustrates this challenge using the video game example from \S \ref{sec:motivating_example}. In this case, the \texttt{Level} object referenced in the \byop{} call site within \texttt{game.py} is defined in a separate file, \texttt{level.py}. Furthermore, \texttt{Level} itself depends on additional nested types declared in yet another file,  \texttt{primitives.py}. Such modular code organization, while common for improving code maintainability, complicates the task of statically gathering all necessary semantic context.

To address these complexities, we introduce MT-IR, a data structure that captures relevant code semantics at each \byop{} call site for prompt generation. MT-IR serves as a bridge between source-level semantics and the prompts for LLM inference. Just as traditional IR is used to generate machine code, MT-IR is used to generate prompts for the model at runtime, where the MT-Runtime relies solely on MT-IR without needing access to the original source code. Figure~\ref{fig:MTIR_for_game} presents an example of MT-IR, capturing code semantics for the source code shown in Figure~\ref{fig:game_code}. The MT-IR is then used to generate the prompt shown in Figure~\ref{fig:genarated prompt}.

The compilation pipeline for generating the MT-IR for each \byop{} call-site involves several key steps:

\subsubsection{Semantic Registry Generation} During the compilation phase of the language, we construct a \textit{semantic registry} using the Abstract Syntax Tree (AST) generated for each code module. This registry serves as a comprehensive repository of all semantic information present in the code, including variable names, function names, and type information. As seen in Figure \ref{fig:game_code}, semantic information from multiple modules (\texttt{game.py}, \texttt{level.py}, and \texttt{primitives.py}) is integrated into a unified data structure. The resulting semantic registry, shown in Figure \ref{fig:semantic_registry}, is a super set of the symbol-table of the language as it contains the definitions as well as the usages of objects, linking usages with definitions.


\begin{wrapfigure}{R}{0.56\textwidth}
    \begin{minipage}{0.56\textwidth}
\begin{small}
\fontsize{6pt}{7pt}\selectfont

\begin{algorithm}[H]
\caption{MT-IR Constructor Algorithm}
\label{algo:mt_ir}
\begin{algorithmic}[1]
\REQUIRE Code Base $C$, Semantic Registry $S$
\ENSURE MT-IR Map $M$
\STATE Initialize MT-IR $M \gets \emptyset$
\FOR{each $\byop{}_i$ call-site in $C$}
    \IF{$\byop{}_i$ is a function-call with name $f_i$}
        \STATE Initialize $M[\byop{}_i] \gets \langle f_i, (T_1,\ldots,T_n) \rightarrow T_r, m, \theta \rangle$ \COMMENT{From equation \ref{eq:function-def-rule}}
    \ELSIF{$\byop{}_i$ is a method-call $mth_i$ in object of class $C_j$}
        \STATE Initialize $M[\byop{}_i] \gets \langle mth_i, C_j, (T_1,\ldots,T_n) \rightarrow T_r, m, \theta \rangle$ \COMMENT{From equation \ref{eq:method-def-rule}}
    \ELSIF{$\byop{}_i$ is an object-initialization of class $C_j$}
        \STATE Initialize $M[\byop{}_i] \gets \langle C_j, (T_1,\ldots,T_k), (T_{k+1},\ldots,T_n), m, \theta \rangle$ \COMMENT{From equation \ref{eq:object-init-rule}}
    \ENDIF
    \STATE \textit{\% Extract Parameter Type Semantics:}
    \FOR{each type $T_j$ in input types of $M[\byop{}_i]$}
            \STATE $type\_def_j \gets \textsc{ExtractTypeDefinition}(T_j, S)$
            \STATE Add $type\_def_j$ to $M[\byop{}_i]$
    \ENDFOR
    
    \STATE \textit{\% Extract Return Type Semantics:}
    \FOR{each type $T_j$ in output/return types of $M[\byop{}_i]$}
        
            \STATE $type\_def_j \gets \textsc{ExtractTypeDefinition}(T_j, S)$
            \STATE Add $type\_def_j$ to $M[\byop{}_i]$
    \ENDFOR
\ENDFOR
\RETURN $M$
\end{algorithmic}
\end{algorithm}

\end{small}
    \end{minipage}
\end{wrapfigure}

\subsubsection{\byop{} Call-site Semantic Extraction}
The created semantic registry is used to traverse and find semantics relevant to each \byop{} call-site. The process begins by extracting semantics from \byop{} call-site. For example, consider the \byop{} call-site in the \texttt{game.py} module, shown in Figure~\ref{fig:game_code} for a the function call. As defined in equation \ref{eq:function-call-rule}, the \byop{} operator calls the MT-Runtime using 
$(f, [v_1, \ldots, v_n], [(T_1, \ldots, T_n), T_r], m, \theta)$ as input information. These are the critical set of semantics relevant to the \byop{} call-site, which needs to be extracted first from the semantic registry as shown in Figure~\ref{fig:MTIR_for_game} according to Algorithm~\ref{algo:mt_ir} line 3-9. Next, the input and output types are checked to determine whether they are primitive types (e.g., integers, floats). If the types are non-primitive, we need to extract their definitions by recursively traversing the semantic registry. Using this traversal, we need to extract the required semantics of the \texttt{Level} object, as illustrated in Figure~\ref{fig:MTIR_for_game}, which depends on \texttt{Map}. In turn, \texttt{Map} relies on \texttt{Wall} and \texttt{Position}. The use-definition links within the semantic registry enable this traversal, ensuring that all types are resolved down to primitive types. This recursive type resolution is signified with "ExtractTypeDefinition" function on Algorithm~\ref{algo:mt_ir}. This approach effectively captures type information for nested types and addressing (\ref{C2}). Algorithm~\ref{algo:mt_ir} presents the complete MT-IR construction process for reproducibility.

The extracted MT-IR can be used to fill in the semantic information in the prompt as shown in Figure~\ref{fig:genarated prompt}, where the highlighted text of the prompt represents all the semantics available in the MT-IR for the given \byop{} call-site.


\begin{figure}[t]
    \begin{subfigure}[t]{\textwidth}
        \centering
        \includegraphics[width=.89\linewidth]{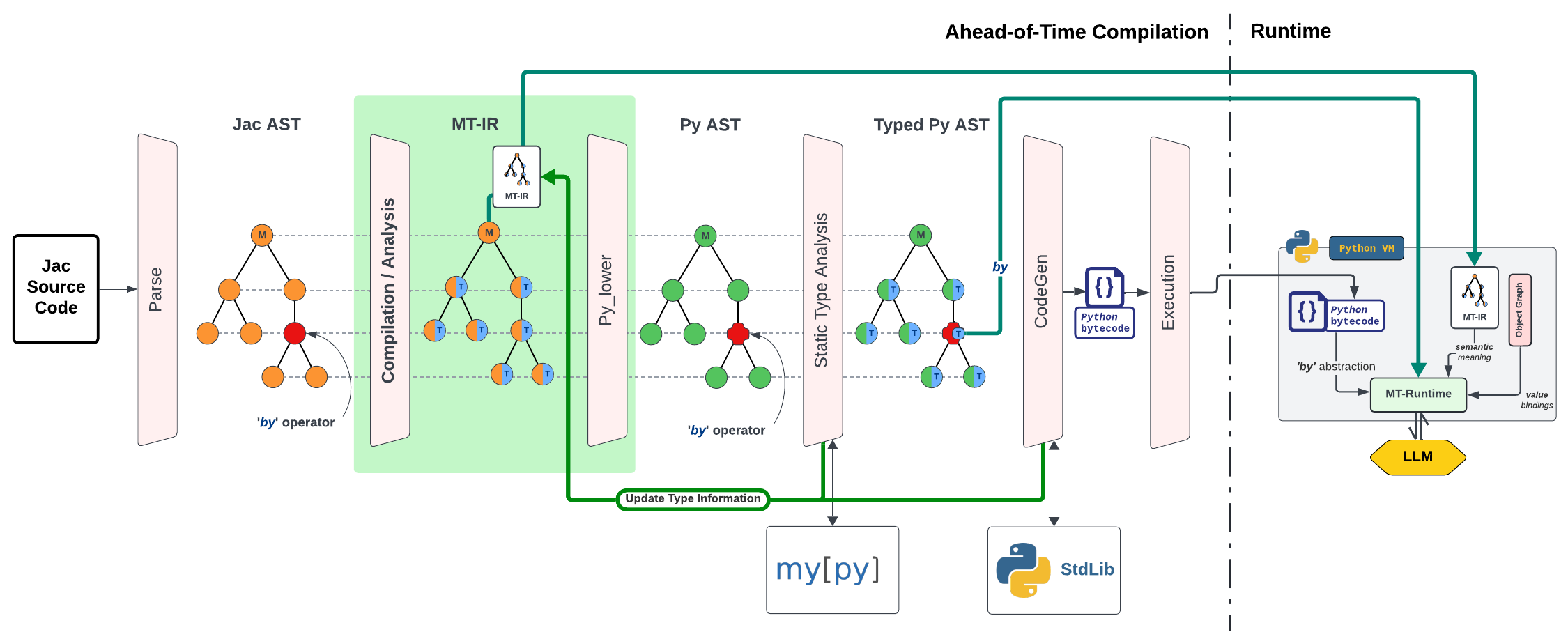}
    \end{subfigure}
\caption{The Jac compiler workflow on which Meaning-Typed compilation is implementation in the Jac programming language.}
\Description{}
\label{fig:jac_compiler}
\end{figure}

\textbf{We implement the \byop{}} language abstraction, MT-IR, and our compiler pass through Jac, a production-grade Python superset language~\cite{mars2023jaseci, jaseci-repo, mars2025objectspatialprogramming, mars2025extendingobjectspatialsemantics}. Typically, when a Python program is executed, the source code is compiled into Python bytecode via ahead-of-time compilation. The bytecode is then interpreted or compiled in a just-in-time fashion at runtime. Jac extends Python in the form of a PyPI package~\cite{jac-pypi}, providing a CLI command, \texttt{jac} that replaces the standard Python compiler by inserting a customized jac compilation phase as in Figure~\ref{fig:jac_compiler}. The resulted python bytecode will then be interpreted using standard \texttt{python3}.

The compilation process for MTP in Jac involves six key steps through the Jac compiler as shown in Figure~\ref{fig:jac_compiler}. Initially, an additional compiler pass generates MT-IR by extracting semantic elements from the AST. This information is then refined through multiple analysis passes. Next, the Jac AST is lowered to a Python AST, where specific code constructs are replaced with MT-Runtime calls (depicted as red nodes). Next, MyPy type checking is performed on the Python AST with resulting type information fed back into the MT-IR to enhance its semantic content. This enriched MT-IR is then registered into the MT-Runtime library, which remains available throughout the program execution. Finally, standard Python bytecode is generated from the modified Python AST, preserving crucial semantic elements in the runtime that would otherwise be lost in conventional compilation in the MT-IR.

\subsection{Meaning-Typed Runtime System}
\label{sec:MT-Runtime}

\begin{wrapfigure}{r}{0.6\textwidth}
    \begin{subfigure}[t]{\linewidth}
        \centering
        \includegraphics[width=1.0\linewidth]{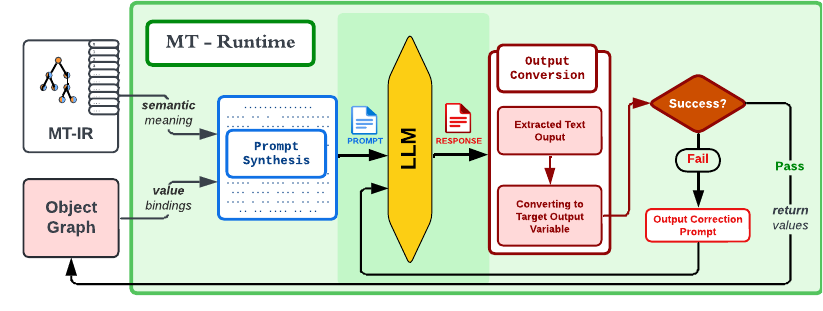}
    \end{subfigure}
\caption{Meaning-typed Runtime System (MT-Runtime).}
\Description{}
\label{fig:MT-Runtime}
\end{wrapfigure}

For AI-Integrated programs, LLM inference is done during the runtime as the prompt requires the semantic information as well as the dynamic values available when the \byop{} is called. This demands a dynamic system which can combine the MT-IR and dynamic value bindings to generate the final prompt for LLM inference.
To this end, we introduce MT-Runtime, a novel automated runtime engine integrated into the language's VM to realize \ref{O3}. It is specifically triggered at the \byop{} call site abstractions, automatically managing the LLM invocation at runtime. MT-Runtime engine effectively addresses the final two challenges mentioned in \S \ref{sec:problem} (\ref{C3} and \ref{C4}).

In this Section, we describe the design of MT-Runtime, which functions through the following conceptual stages.

\subsubsection{Prompt Synthesis}

Upon the triggering of a~\byop{} call, MT-Runtime first retrieves the MT-IR which is relevant to that particular \byop{} call-site and uses its information to fill a generalized prompt template. Based on the location of the~\byop{} call-site, MT-Runtime constructs a prompt that incorporates static semantics, dynamic variable values relevant to the call-site, and the expected output type. As the MT-Runtime resides on the language VM  as shown in Figure \ref{fig:jac_compiler}, it has access to the variable values at runtime. In Figure \ref{fig:MT-Runtime}, this can be seen where the semantic information is fetched from the MT-IR while the variable values are fetched from the languages object graph. The synthesized prompt can be seen in Figure~\ref{fig:genarated prompt}. Here, the function name determines the `Action' in the prompt, representing the intended task, while type information and nested types are included under `Type\_Explanations' in the prompt. Additionally, the inclusion of the output type and other schema information in the prompt forces the LLM to generate the output as per the defined schema, making it easier to convert in the type conversion stage. This stage address the challenge of generating meaningful but concise prompts discussed in \S\ref{sec:problem} as \ref{C3}.

\subsubsection{Conversion from LLM Output to Target Output Variable}

MT-Runtime queries the LLM with the dynamically synthesized input prompt and receives an output. The next step is to parse the LLM’s textual output and convert it into the target output variable addressing \ref{C4}.

During the prompt synthesis phase, the prompt instructs the model to generate output following the predefined Python object schema. MT-Runtime leverages Python's built-in \texttt{ast.literal\_eval()} to evaluate the output text and construct the corresponding variable instance. For example, as in Figure~\ref{code: object-init}, the variable \texttt{einstein} is of the custom type \texttt{Person} with attributes \texttt{name: str} and \texttt{dob: str}. In this case, MT-Runtime instructs the LLM to generate an output as \texttt{Person(name=``Albert\ Einstein",\ dob=``03/14/1879")}, which can be directly evaluated into a valid \texttt{Person}-typed object.

When MT-Runtime encounters an output that cannot be converted into the target variable type, it constructs a revised prompt that attempts to rectify the error as illustrated in Figure \ref{fig:MT-Runtime}. This revised prompt includes details of the current incorrect output and the expected type, guiding the model toward the correct format. MT-Runtime queries the LLM with this revised input and attempts to parse its output again. These retry attempts continue until either the type conversion succeeds or the maximum number of retries specified by the developer is reached. If the LLM fails to generate a correct output at the end, the MT-Runtime raises an exception with a type error. While errors are handled elegantly through this approach, the revised prompt is mostly short and concise since it only aims to correct the type mismatches. This offers an efficient solution for \ref{C4}, handling errors gracefully. 

\section{Evaluation}
\label{sec:Eval}

\begin{table}
\centering
\caption{Benchmark Suite Descriptions and Correctness Criteria.}
\resizebox{\textwidth}{!}{%
\begin{tabular}{|p{0.2\textwidth}|p{0.7\textwidth}|p{0.7\textwidth}|}
\hline
\textbf{Benchmarks} & \textbf{Task Description} & \textbf{Correctness Criteria} \\ \hline
\textbf{Math Problem} & Given a grade school math question an integer answer is returned. \cite{khattab2023dspy} & Answer matches the correct integer solution. \\ \hline
\textbf{Translation}  & Translates a given text from English to French. \cite{Beurer2023LMQL} & Translation preserves meaning and fluency. \\ \hline
\textbf{Essay Reviewer}  & Reviews an essay and assigns a letter grade based on the quality of the essay. \cite{zheng2023SGLang} & Grade is appropriate for essay content and structure. \\ \hline
\textbf{Joke Generator}  & Generates jokes based on a given topic, e.g., jokes with punchline. \cite{Beurer2023LMQL} & Joke is coherent, on-topic, and has a valid punchline. \\ \hline
\textbf{Expert Answer}  & Given a question, predicts the expert profession best suited to answer and generates an answer. \cite{zheng2023SGLang} & Expert domain is appropriate and the answer is relevant and accurate. \\ \hline
\textbf{Odd Word Out}  & Identifies the odd word out in a list of words along with a reason for why it is the odd word out. \cite{Beurer2023LMQL} & Correct word identified with valid justification. \\ \hline
\textbf{MCQ Reasoning}  & Answers multiple-choice questions by providing a reason for the chosen answer. \cite{Beurer2023LMQL} & Correct choice selected with logically sound reasoning. \\ \hline
\textbf{Personality Finder}  & Given a public figure's name, predicts their personality type. & Predicted personality matches known or inferred type. \\ \hline
\textbf{Template}  & Given a template object and a set of variables, fills in the template with the variables. \cite{Beurer2023LMQL} & All template fields filled correctly with appropriate values. \\ \hline
\textbf{Text to Type}  & Given a text input and a structured data type, extracts the structured data into the structured data type. \cite{Beurer2023LMQL} & Extracted output matches the target structured format exactly. \\ \hline
\textbf{Taskman} & Given a list of tasks, assigns priority ranks and estimates completion times for each task. & Task rankings and estimates are reasonable and well justified. \\ \hline
\textbf{Level Generator} & Generates new video game-levels based on previous levels. & Two playable levels are generated, with the second level being harder than the first based on complexity and enemy count. \\ \hline
\textbf{Wikipedia} & Given a question and an external tool (e.g., Wikipedia), retrieves the relevant answer using ReAct methodology. \cite{Beurer2023LMQL} & Answer is factually correct and supported by retrieved evidence. \\ \hline
\end{tabular}
}
\label{tab:benchmarks}
\end{table}

In this section, we present our evaluation of \systemname{} in comparison to prior work (LMQL~\cite{Beurer2023LMQL} and DSPy~\cite{khattab2023dspy}) aimed at reducing the complexities associated with prompt engineering. With reduced developer effort, we strive to ensure that \systemname{} maintains accuracy without increasing token usage with respective to prior works. Additionally, our lightweight runtime system is designed to minimize both cost and runtime overhead. Furthermore, we assess the robustness of MTP in scenarios where poor coding practices diminish the semantic richness of the code. Our evaluation aims to answer the following research questions.

\begin{enumerate}[label=\textbf{RQ$_\arabic*$}]

    \item \label{RQ1} - \textbf{\textit{How effectively does \systemname{} reduce development complexity for AI-Integration?}}  

    \item \label{RQ2} - \textbf{\textit{How does the accuracy of the \systemname{} compare to prior work?}}

    \item \label{RQ3} - \textbf{\textit{What are the trade-offs for token usage, cost, and runtime in \systemname{}?}}

    \item \label{RQ4} - \textbf{\textit{How resilient is MTP to sub-optimal coding practices?}}
    
\end{enumerate}

To evaluate these research questions, we introduce, (i) two case-studies, (ii) a user study and (iii) a suite of benchmarks. 


\begin{enumerate}[label=\textbf{(\arabic*.)}, nosep]
       \item \textbf{Case Studies} (\ref{RQ1}). We first qualitatively evaluate the reduction of complexities for AI-Integration using \systemname{} (\S~\ref{sec:case}).

        \item \textbf{Lines of code (LOC)} (\ref{RQ1}). In addition to the user study, we evaluate \systemname{} for its effectiveness in reducing programming complexity through a quantitative study measuring lines of code across a comprehensive suite of benchmarks (\S~\ref{sec:LOC}).
    
       \item \textbf{User study} (\ref{RQ1}). We conducted a user study to evaluate \systemname{}'s effectiveness in reducing complexity and increasing developer's productivity (\S~\ref{sec:user}).
       
        \item \textbf{Accuracy and Trend across LLM evolution} (\ref{RQ2}). We evaluate the program accuracy when using \systemname{} versus prior work across 10 various LLMs (\S~\ref{sec:accracy}).
    
       \item \textbf{Token Usage, Cost and Runtime} (\ref{RQ3}). We measured these performance metrics for \systemname{} compared to DSPy (\S~\ref{sec:token}).
    
       \item \textbf{Sensitivity To Coding Practices} (\ref{RQ4}). Here we analyze how the \systemname{} paradigm can be affected through bad coding practices using the game level generation benchmark. (\S~\ref{sec:sensitivity})
    
\end{enumerate}

\paragraph{\textbf{Experimental Setup}}
We implemented \systemname{} using Jaseci~\cite{mars2023jaseci,jaseci-repo,mars2025objectspatialprogramming}, an open-source research infrastructure that offers full control over language semantics, compilation processes, and runtime execution, all built as a superset of Python.

We use OpenAI APIs for experiments with GPT models. We host llama models on a local server with  Nvidia 3090 GPU with 24GB VRAM and 64GB RAM. We instrument the source code to profile token usage. Python cProfile\cite{python_cProfile} is used to measure program runtime. Our benchmark suite consists of problems representative of applications with AI-Integration. To construct the suite, we use the entire set of benchmarks and examples from prior work \cite{Beurer2023LMQL}, supplemented with all relevant benchmarks in other work \cite{zheng2023SGLang, khattab2023dspy} (Table \ref{tab:benchmarks}).

\subsection{\textbf{A Case Study : Dynamic Level Generation in Video Game Development}}
\label{sec:case}

\begin{wrapfigure}{r}{0.55\textwidth} 
    \centering
    \includegraphics[width=0.8\linewidth]{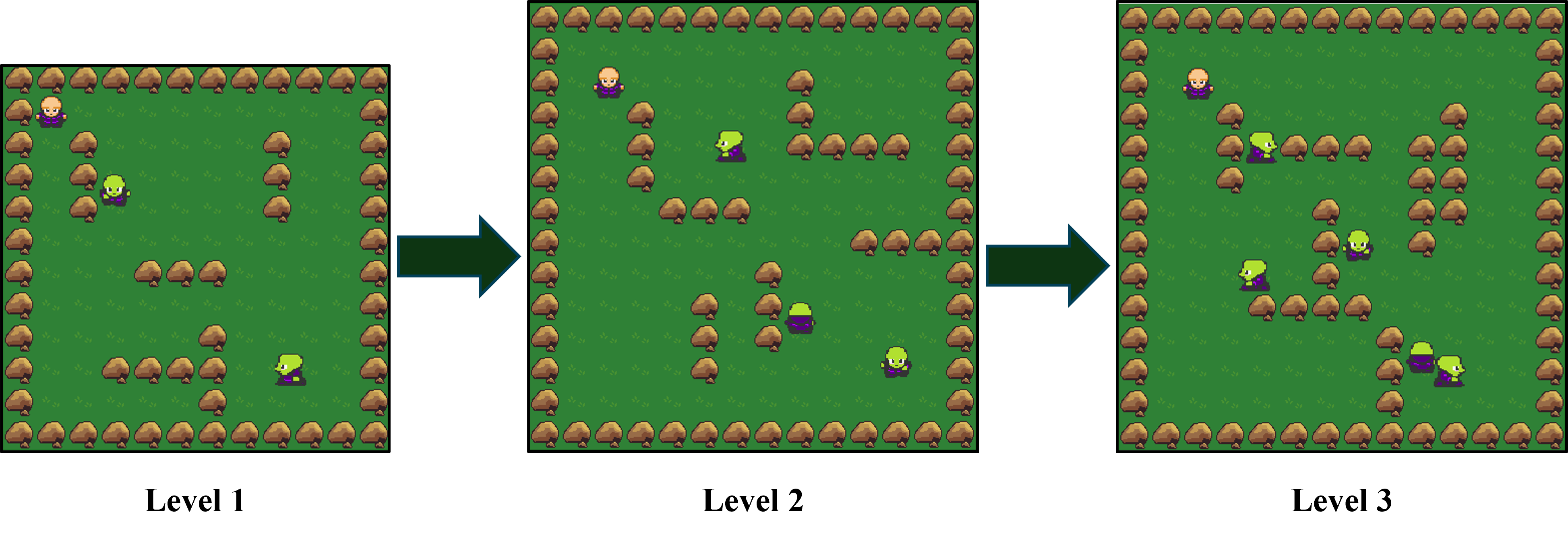}
    \Description{Three consecutive game levels generated by an LLM utilizing \systemname{} feature. Each level designed to be harder than the previous level.}
    \caption{A sequence of three game levels dynamically generated by an LLM at runtime.}
    \label{fig:rpg_levels}
     \vspace{0.9cm}
\end{wrapfigure}

\S \ref{sec:motivating_example} presented an example use-case of AI-Integration for automatic level generation of a video game. This video game is a real program developed using an online tutorial~\cite{pygame2021tutorial} using the pygame~\cite{mcgugan2007beginning} library. Figure \ref{fig:rpg_levels} represents three consecutive levels generated by an LLM through AI-Integration.

\begin{figure}
    \centering
    \begin{minipage}{0.49\textwidth}
        \input{code/game_lmql}
        (a) AI-Integrated Game Levels : LMQL
    \end{minipage}
    \hfill
    \begin{minipage}{0.49\textwidth}
        \begin{minipage}{\textwidth}
            \input{code/game_dspy}
            (b) AI-Integrated Game Levels : DSPy
        \end{minipage}
        \begin{minipage}{\textwidth}
            \vspace{4.98cm}
        \end{minipage}
        \begin{minipage}{\textwidth}
            \input{code/MTP_game}
            (c) AI-Integrated Game Levels : MTP
        \end{minipage}
    \end{minipage}
    \caption{Implementation of the video game usecase using (a) LMQL and (b) DSPY frameworks. (c) This gives a preview how the same task can be achieved using the MTP paradigm introduced in the paper.}
    \label{fig:game prior}
\end{figure}

First we examine how this can be achieved using prior work, LMQL~\cite{Beurer2023LMQL} and DSPy~\cite{khattab2023dspy} which aims to simplify prompt engineering. Figure~\ref{fig:game prior} includes the two code snippets for implementation from each framework.

\paragraph{\textbf{LMQL~\cite{Beurer2023LMQL}}} Language Model Query Language introduces a SQL-like query language for LLM prompting with a focus on constraining model outputs. This framework enables developers to enforce structured responses through predefined rules and output schemas. As shown in Figure \ref{fig:game prior}a, LMQL provides a structured approach to our video game level progression use case. While the core prompt (lines 8-34) contains similar instructions to traditional prompt engineering methods in Figure \ref{fig:video_game_PE}, LMQL's key advantage lies in its explicit output format definition (lines 22-33), which specifies expected data types and formats for each attribute.
During inference, LMQL operates iteratively, generating one output field at a time using uppercase text placeholders to mark generation points. Each successfully generated value is injected back into the prompt, repeating until all required values are obtained. While LMQL improves control over LLM-generated outputs, it does not fully abstract away prompt-engineering, as developers still need to construct base prompts manually. This limits its effectiveness in simplifying AI-Integrated application development.

\paragraph{\textbf{DSPy~\cite{khattab2023dspy}}} This developer-friendly framework offers a more structured approach than LMQL for AI-Integration. For simple tasks like question answering, developers use \textit{dspy.Signature} to define structured inputs and outputs. However, complex tasks requiring typed outputs still necessitate manual code modifications. Figure \ref{fig:game prior}b shows our DSPy implementation where lines 30–34 define a \texttt{dspy.Signature} with task description and explicit input/output fields.
A key challenge is DSPy's inability to infer what a \texttt{Level} object represents without parsing the full codebase. As a workaround, developers must explicitly annotate class definitions to inherit from Pydantic's \texttt{BaseModel} (lines 7, 11, 15, and 21), with attributes defined as \texttt{Field} objects containing descriptions. While DSPy improves usability over generic prompt engineering and LMQL, its Pydantic reliance introduces a steep learning curve and forces developers to modify existing code accordingly.

\textbf{In contrast, our \systemname{} implementation} introduces a simple abstraction that reduces complexity for developers. \systemname{} achieves the same functionality with just a few lines of code additions and modifications (Figure \ref{fig:game prior}c), compared to the 58 lines of code modifications required for LMQL and 36 lines for DSPy integration (here \textit{LoC modified} metric signifies the developer complexity of integrating an LLM through the respective framework when adding a new functionality, further elaborated in \S~\ref{sec:LOC}). \systemname{} maintains the same function signature and leverages the semantics of this signature along with the data structures in Figure \ref{fig:video_game_datastructures} to automate prompt generation. MT-runtime engine described in \S \ref{sec:MT-Runtime} automates LLM output parsing and conversion to custom-typed variables, eliminating additional developer effort.

When comparing all three implementations, only LMQL includes map generation rules and instructions in its prompts, whereas DSPy and MTP do not, as they do not require manual prompt engineering.  However, despite the absence of manual prompts, MTP consistently maintains higher accuracy than LMQL, as we show later in Figure~\ref{fig:accuracy_correctness} and \S~\ref{sec:accracy}. We hypothesize that this is due to the fact that MTP uses code semantics to infer the nature of the task. In this case, the source code semantics, such as the variable names and values associated to \texttt{previous\_levels} and \texttt{difficulty}, enable the LLM to infer the need for increasing difficulty across game levels, achieving high accuracy. It is worth noting that additional specific rules such as “width/levels increase by 10\%” can also be passed as arguments to the \byop{} call-site, allowing MTP to consistently apply them during level generation.

\begin{figure}[t]
    \centering

    \begin{minipage}{0.57\textwidth}
        \centering
        \includegraphics[width=\linewidth]{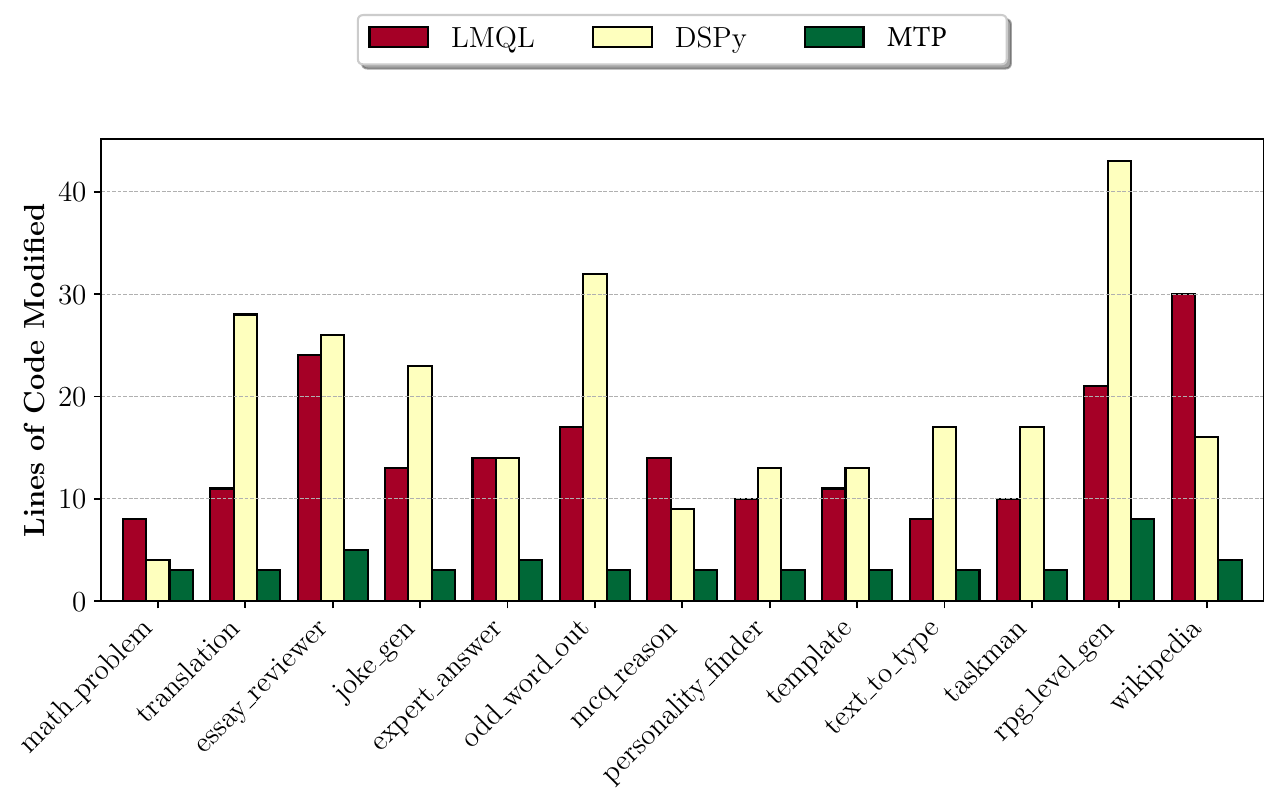}
        \caption{Lines of Code Modifications required for LLM integration into benchmark programs with for implementing a new feature.}
        \label{fig:LoC_modified_dspy_vs_mtllm}
    \end{minipage}
    \hfill
    \begin{minipage}{0.350\textwidth}
        \centering
        \captionof{table}{Lines of Code Modified reduction achieved by \systemname{} Compared to DSPy and LMQL for LLM integration.}
        \resizebox{\linewidth}{!}{%
        \begin{tabular}{|c|c|c|}
        \hline
        \textbf{Problem} &\multicolumn{2}{c|}{\textbf{LoC Modified}}\\
        & \multicolumn{2}{c|}{\textbf{for MTP}} \\
        \cline{2-3}
        &  vs. LMQL & vs. DSPy \\
        \hline
        Math Problem & \down{2.7} & \down{1.3}\\
        \hline
        Translation &  \down{3.7} & \down{9.3}\\
        \hline
        Essay Reviewer & \down{4.8} & \down{5.2}\\
        \hline
        Joke Generator &  \down{4.3} & \down{7.7}\\
        \hline
        Expert Answer & \down{3.5} & \down{3.5}\\
        \hline
        Odd Word Out & \down{5.7} & \down{10.7}\\
        \hline
        MCQ Reasoning & \down{4.7} & \down{3.0}\\
        \hline
        Personality Finder & \down{3.3} & \down{4.3}\\
        \hline
        Template & \down{3.7} & \down{4.3}\\
        \hline
        Text to Type & \down{2.7} & \down{5.7}\\
        \hline
        Taskman & \down{3.3} & \down{5.7}\\
        \hline
        RPG Level Generator & \down{2.3} & \down{4.8}\\
        \hline
        Wikipedia (ReAct) &  \down{7.5} & \down{4.0}\\
        \hline
        \hline
        \textbf{Average} & \textbf{\down{4.0}} & \textbf{\down{5.3}} \\
        \hline
        \hline
        \end{tabular}
        }
        \label{tab:llm_integration}
    \end{minipage}
\end{figure}

\subsection{\textbf{Lines of Code (LOC) Modified Comparison}}
\label{sec:LOC}

We evaluate \systemname{} for its effectiveness in reducing programming complexity when integrating an LLM into traditional programs that do not use any LLM features. We conduct a quantitative study measuring \textit{lines of code (LOC) modified} across our benchmark suite (Table~\ref{tab:benchmarks}), where each benchmark was implemented using all three frameworks: LMQL, DSPy, and \systemname{}. \textbf{\textit{LOC modified} for LLM integration counts the number of lines that need to be added or modified when a developer integrates a new LLM feature into an existing codebase.} LOC-based metrics have also been used in prior work to assess integration effort in language model programming frameworks~\cite{Beurer2023LMQL}. Figure~\ref{fig:LoC_modified_dspy_vs_mtllm} presents the LOC modifications required for AI-Integration across benchmarks, starting from a non-LLM integrated baseline.  The results show that \systemname{} requires the minimum code modifications to integrate an LLM into traditional programming. DSPy and LMQL require much larger LOC across all benchmarks. This difference is due to the simple yet intuitive abstraction introduced in \systemname{}.
 
Table~\ref{tab:llm_integration} presents the \textit{LOC Modified} reduction factors for MTP, illustrating how it simplifies LLM feature integration with reduced implementation effort relative to LMQL and DSPy. It should be noted that these factors were derived form the LoC modified represented on Figure \ref{fig:LoC_modified_dspy_vs_mtllm}. This large reduction is because DSPy requires tedious type annotations and LMQL requires manual programmatic prompt engineering. In contrast, \systemname{} requires minimum code change for LLM integration by simply leveraging the \byop{} operator and hiding all complexity.
\subsection{\textbf{User Study Evaluating MTP against LMQL and DSPy}}
\label{sec:user}

To quantify the degree of reduction in developer complexity and effort achieved through \systemname{}, compared to prior work, we conducted a user study consisting of 20 software developers.

\begin{table}[t]
\caption{Questions asked in the user study questionnaire, grouped under five criteria: Ease of Use, Clarity of Documentation, Learning Curve, Efficiency of Problem-Solving, and Overall Satisfaction.}
\begin{center}
\renewcommand{\arraystretch}{1.3
}
\resizebox{0.8\columnwidth}{!}{%
\begin{tabular}{|>{\centering\arraybackslash}m{2.3cm}|>{\centering\arraybackslash}m{2.5cm}|m{13cm}|}
  \hline
  \textbf{Category} & \textbf{Question \#} & \textbf{Statement} \\
  \hline
  \hline
  \multirow{4}{2.3cm}{\textit{Ease of use}} & Q1 & How easy was it to set up and start using the framework? \\
  \cline{2-3}
  & Q2 & How intuitive do you find the syntax and structure of the framework? \\
  \cline{2-3}
  & Q3 & How would you rate the ease of performing common tasks with the framework? \\
  \cline{2-3}
  & Q4 & How quickly will be able to integrate the framework into your existing projects? \\
  \hline
  \hline
  \multirow{4}{2.3cm}{\textit{Clarity of Documentation}} & Q5 & How clear and understandable do you find the official documentation? \\
  \cline{2-3}
  & Q6 & How well is the documentation structured to find information quickly? \\
  \cline{2-3}
  & Q7 & How helpful are the provided examples and tutorials in understanding the framework? \\
  \cline{2-3}
  & Q8 & How complete and detailed are the code examples in the documentation? \\
  \hline
  \hline
  \multirow{3}{2.3cm}{\textit{Learning Curve}} & Q9 & How long did it take you to feel comfortable with the basic features of the framework? \\
  \cline{2-3}
  & Q10 & How easy was it to learn and implement advanced features of the framework? \\
  \cline{2-3}
  & Q11 & How well does the framework support users in learning and utilizing advanced concepts? \\
  \hline
  \hline
  \multirow{2}{2.3cm}{\textit{Efficiency of Problem-Solving}} & Q12 & How efficient is the framework in solving problems compared to others you have used? \\
  \cline{2-3}
  & Q13 & How easy was it to learn and implement advanced features of the framework? \\
  \hline
  \hline
  \multirow{4}{2.3cm}{\textit{Overall Satisfaction}} & Q14 & Overall, how satisfied are you with the framework? \\
  \cline{2-3}
  & Q15 & Would you recommend this framework to others? \\
  \cline{2-3}
  & Q16 & How likely are you to continue using this framework in your future projects? \\
  \cline{2-3}
  & Q17 & What are the main reasons for your rating in the previous question? \\
  \hline
\end{tabular}
}
\Description{}
\end{center}
\label{tab:user_study_questions}
\end{table}

\subsubsection{User Study Protocol}

\begin{wrapfigure}{r}{0.45\textwidth}
    \centering
    \includegraphics[width=\linewidth]{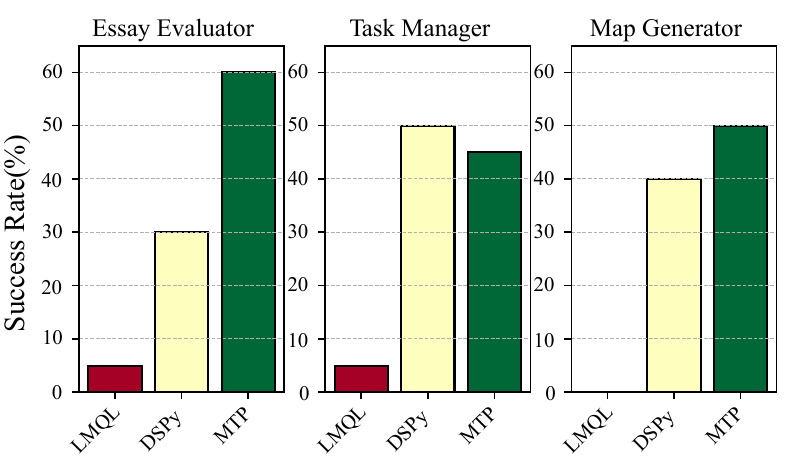}
    \caption{User study success rates across LMQL, DSPy, and~\systemname{}.}
    \label{fig:user-study-success-rate}
\end{wrapfigure}

We recruited 20 undergraduate students from a Natural Language Processing class who had Python coding experience but minimal LLM-integrated development background (limited to basic prompt engineering). This allowed us to evaluate developer complexity from a newcomer's perspective, as participants lacked prior experience with the baseline frameworks or \systemname{}. Participants were given one day to familiarize themselves with \systemname{}, DSPy, and LMQL through standard documentation, tutorials, and online resources, then tasked with implementing three progressively challenging problems using all three frameworks: \emph{Essay Evaluator (Easy)}, \emph{Task Manager (Medium)}, and \emph{Game Level Generator (Hard)}. Each task was selected from our benchmark suite to represent varying complexity levels, with 90 minutes allocated per implementation. Upon completion, participants provided feedback on each framework's usability and effectiveness through a comprehensive questionnaire. To ensure the integrity of the study, responses were collected anonymously, participants were compensated for their time, and the sole objective disclosed to them was the evaluation of three different LLM integration frameworks.

\subsubsection{Success Rate}

Figure \ref{fig:user-study-success-rate} presents the success rates across three programming tasks using DSPy, LMQL, and \systemname{}. Success is defined as when the implementation a participate developed generates accurate outputs across test inputs. Overall, implementation utilizing \systemname{} achieved the highest success rate in 2 out of the 3 tasks.  It also performs the most consistently across all tasks. These results demonstrate \systemname{}'s effectiveness in enabling programmers to leverage LLM capabilities intuitively. In contrast, we observe that participants found using LMQL to be the most challenging, with the lowest success rates. Qualitative feedback from participants indicated that this is because LMQL requires developers to manually craft prompts while DSPy and \systemname{} automate the process. For the most challenging problem, \emph{Game Level Generator (Hard)}, developers achieved zero success with LMQL, highlighting the difficulty to manually craft prompt and than manually interpret output to convert to a complex object type.

\begin{figure}[t]
    \centering
    \includegraphics[width=\linewidth]{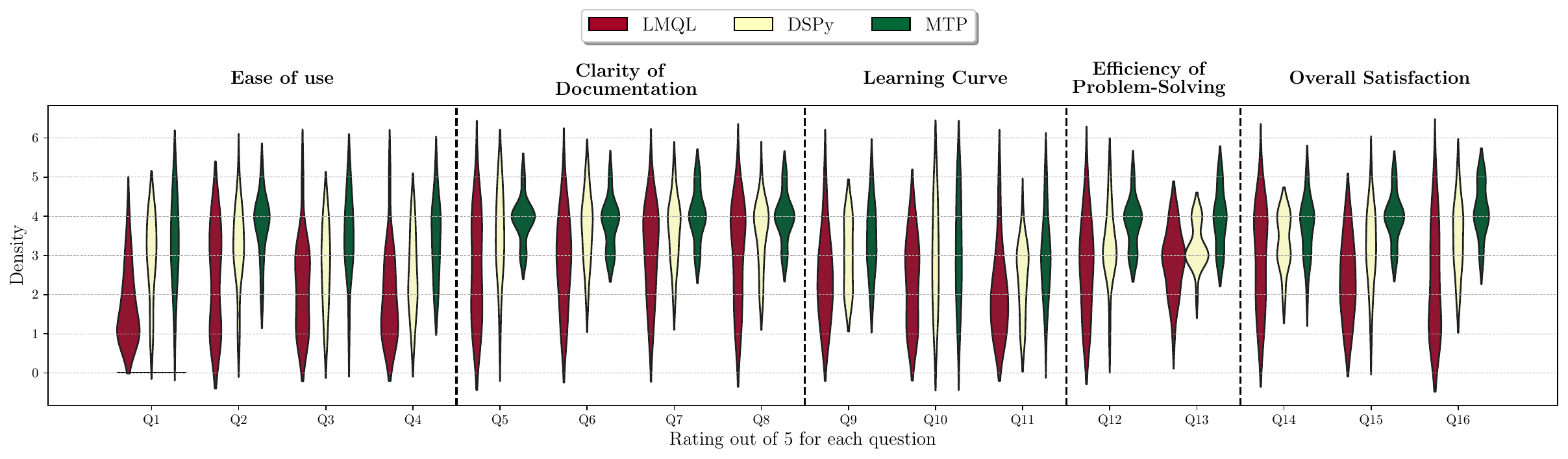}
    \Description{}
    \caption{Violin plot showing the distribution of scores given by participants for each framework across the five criteria: Ease of Use, Clarity of Documentation, Learning Curve, Efficiency of Problem-Solving, and Overall Satisfaction.}
    \label{fig:user_study_dist}
\end{figure}

\begin{wrapfigure}{r}{0.45\textwidth}
    \vspace{-0.5cm}
    \centering
    \includegraphics[width=\linewidth]{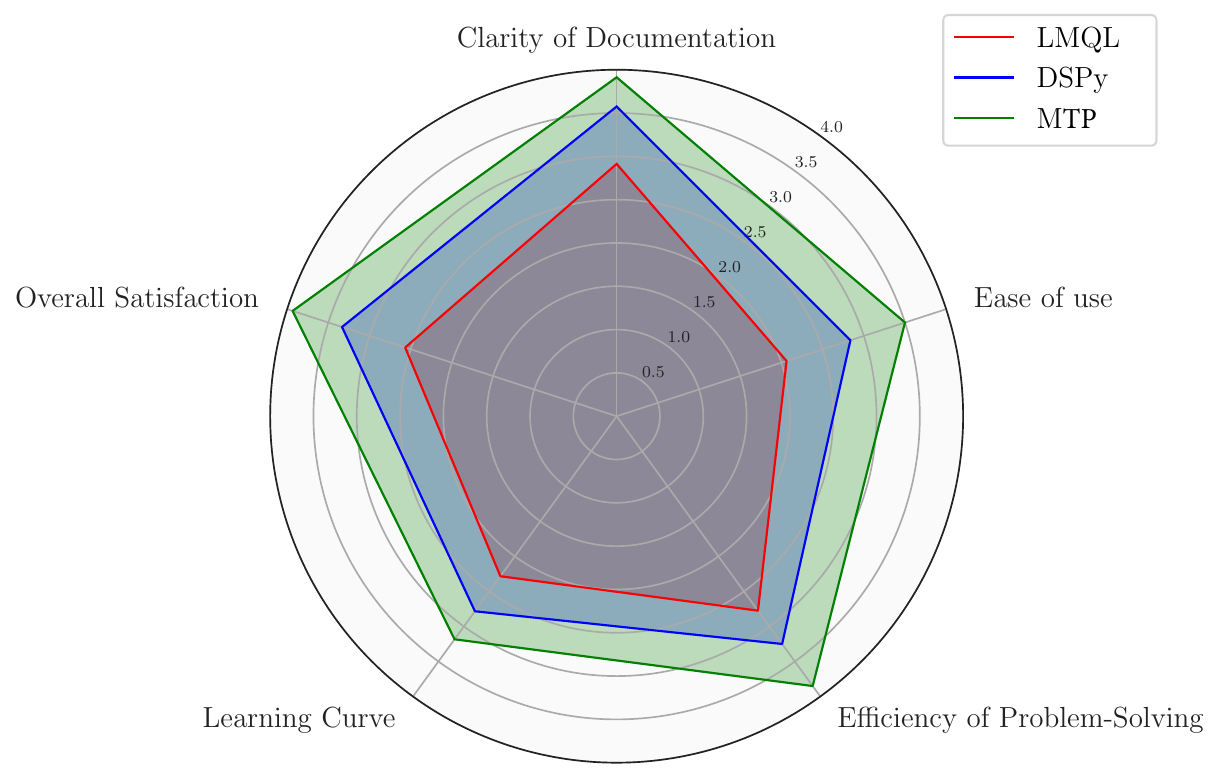}
    \Description{User evaluation of LMQL, DSPy and \systemname{} on five usability criteria.}
    \caption{User evaluation of LMQL, DSPy and \systemname{} on five usability criteria}
    \label{fig:usability-radar}
\end{wrapfigure}

\subsubsection{User Study Feedback}

 Our questionnaire consists of 17 questions grouped under five criteria: Ease of Use, Clarity of Documentation, Learning Curve, Efficiency of Problem-Solving, and Overall Satisfaction (Table~\ref{tab:user_study_questions}). 
The average user scores for each framework across the five criteria are shown in Figure \ref{fig:usability-radar}. \systemname{} was consistently rated the better framework compared to LMQL and DSPy. DSPy generally scored in the mid-range, while LMQL scored the lowest across all criteria, suggesting it introduced the most complexity for programmers. 

The Violin plot in Figure \ref{fig:user_study_dist} shows the distribution of scores given by participants for each framework. Across all criteria, \systemname{} exhibits higher density around the upper rating regions (higher scores) compared to LMQL and DSPy, indicating a generally more favorable rating. Notably, for categories such as Ease of use, efficiency of problem solving, \systemname{} achieved higher density peaks near the higher scores, reflecting higher perceived developer productivity and lower complexity of \systemname{} compared to the other frameworks.

Open-ended participant feedback further highlighted \systemname{}'s strengths, with comments such as \textit{"Learning LMQL was challenging, and it ultimately proved inadequate for accomplishing the required tasks."}, \textit{"MTP provided a much simpler and more precise solution, making it the most efficient tool for the tasks at hand."} and \textit{"MTP code is shorter than the other two frameworks... My favorite feature is the automatic filling of object attributes using by llm.'} \textit{ conversions between data types can be done easily."}. 

Using the results of the case study, user study and LoC reduction evaluation  we can conclude that the abstraction offered through MTP reduces complexity for developers streamlining model-integration compared to other available solutions.

\subsection{Program Accuracy}
\label{sec:accracy}

\begin{figure*}[t] 
        \centering
        \includegraphics[width=0.98\linewidth]{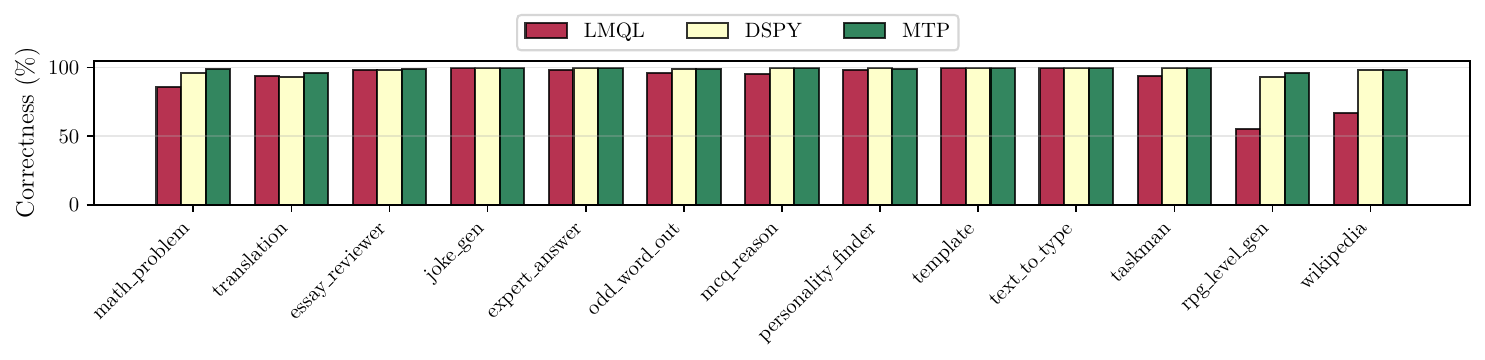}
            \caption{Accuracy of each benchmark relative to the correctness criteria defined for each benchmark, for 100 trails.}
            \label{fig:accuracy_correctness}
        
\end{figure*}
In this section, we evaluate whether \systemname{}'s complexity reduction comes at a cost of accuracy loss and investigate the impact of LLM evolution on program accuracy when using \systemname{} compared to previous work.

\subsubsection{Correctness across benchmark suite}

We evaluate our system using the benchmark suite outlined in Table~\ref{tab:benchmarks}, employing GPT-4o. Each benchmark includes task-specific correctness criteria, also detailed in Table~\ref{tab:benchmarks}. Our evaluation consists of 100 runs per benchmark, derived from 20 distinct test inputs, each evaluated 5 times. As many benchmarks involve generative tasks without a single ground truth answer, we adopt different evaluation methodologies based on task type. These, summarized in Table~\ref{tab:eval_methods}, include GPT4Score~\cite{fu-etal-2024-gptscore} and human evaluation for open-ended tasks. Across all methods, we apply a binary correctness standard: only fully correct outputs are marked as correct, with partially correct answers considered incorrect. For each framework, we report accuracy as the percentage of correct outputs across the 100 trials for each benchmark.

\begin{table}[h]
\centering
\caption{Evaluation methodologies for benchmark correctness assessment}
\label{tab:eval_methods}
\resizebox{0.98\columnwidth}{!}{%
\begin{tabular}{|p{3cm}|p{6.5cm}|p{12cm}|}
\hline
\textbf{Method} & \textbf{Description} & \textbf{Applied Benchmarks} \\
\hline
ExactMatch & Direct equality with ground truth & \texttt{text\_to\_type}, \texttt{mcq\_reason}, \texttt{math\_problem} \\
\hline
GPTScore~\cite{fu-etal-2024-gptscore} & GPT model as judge~\cite{zheng2023judging, fu-etal-2024-gptscore} evaluating compliance with criteria & \texttt{essay\_reviewer}, \texttt{expert\_answer}, \texttt{wikipedia}, \texttt{template}, \texttt{taskman}, \texttt{translation}, \texttt{personality\_finder}, \texttt{joke\_gen}, \texttt{odd\_word\_out} \\
\hline
Human Evaluation & Human evaluation based on defined criteria, conducted by members of the lab cohort. & \texttt{rpg\_level\_gen} \\
\hline
\end{tabular}
}
\end{table}

Figure~\ref{fig:accuracy_correctness} shows that all frameworks achieve above 50\% correctness across all benchmarks. It should be noted that the correctness of LMQL is directly tied to how well developers manually craft prompts, unlike the automated prompt generation in MTP and DSPy. This manual effort increases complexity and reduces developer productivity, as further demonstrated in our user study (\S~\ref{sec:user}). Therefore, LMQL implementations are used as provided in the source material when available, or otherwise faithfully implemented. LMQL demonstrates lower accuracy than DSPy and \systemname{} in 10 out of 13 benchmarks. DSPy and \systemname{} show nearly identical performance, with \systemname{} outperforming DSPy in 9 out of 13 benchmarks. The average correctness across benchmarks is 90.85\% for LMQL, 98.23\% for DSPy, and 98.92\% for \systemname{}, indicating that \systemname{} maintains competitive correctness while significantly reducing developer complexities as shown in \S~\ref{sec:case}, \S~\ref{sec:LOC} and \S~\ref{sec:user}.

\begin{figure*}[t] 
  
  \begin{minipage}[c]{0.35\linewidth}
    \centering
    \includegraphics[width=\linewidth]{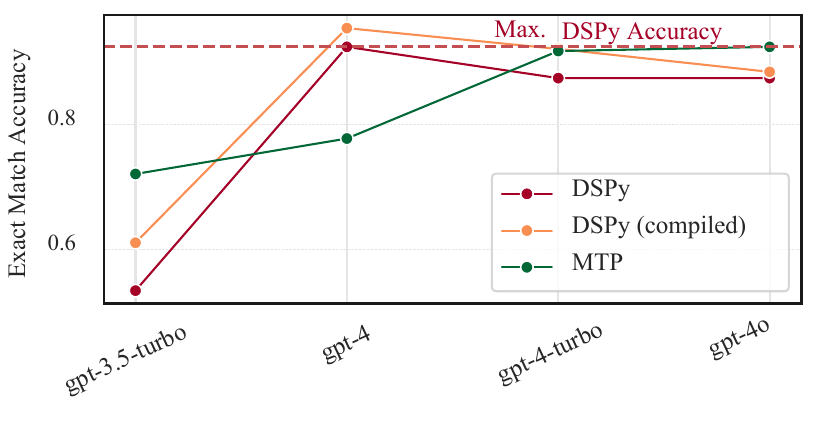}
            \caption{Accuracy on GSM8K across OpenAI models.}
            \label{fig:gsm_exactmatchaccuracy_openai_mtllm}
  \end{minipage}
  \hfill 
  \begin{minipage}[c]{0.35\linewidth}
    \centering
    \includegraphics[width=\linewidth]{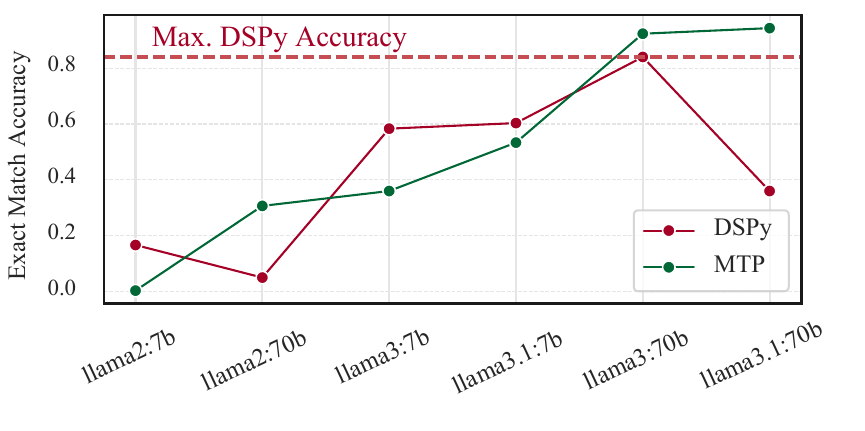}
    \caption{Accuracy on GSM8K across Llama models.}
    \label{fig:gsm_exactmatchaccuracy_llama_mtllm}
  \end{minipage}
  \hfill 
  \begin{minipage}[c]{0.27\linewidth}
    \centering
    \includegraphics[width=\linewidth]{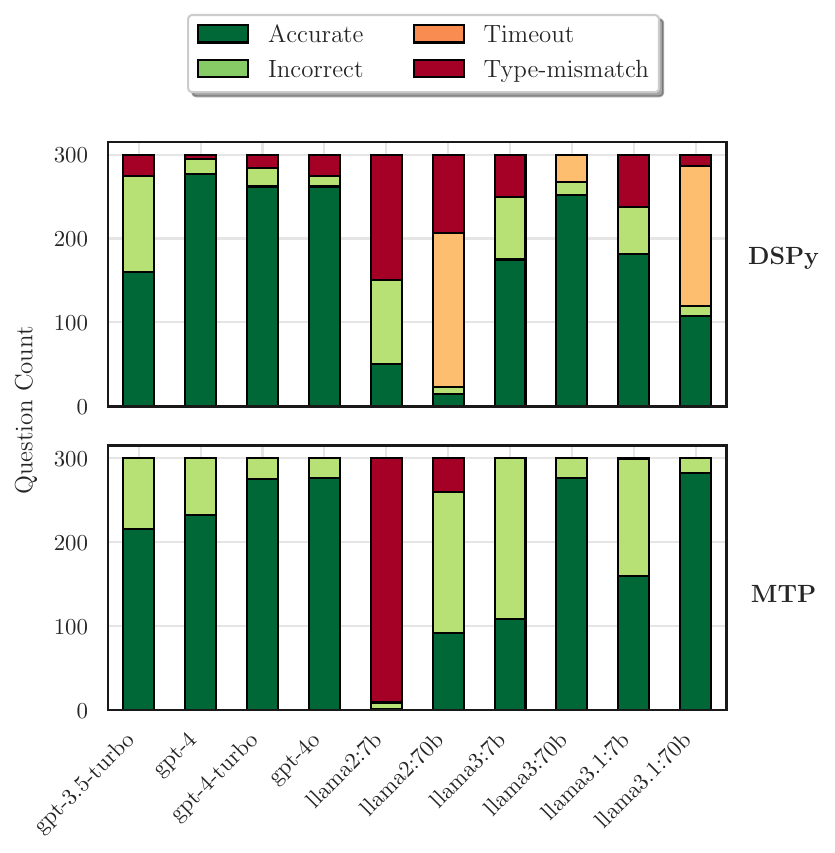}
    \caption{Results breakdown for DSPy and \systemname{} on GSM8K. (Timeout set at 120s).}
    \label{fig:gsm_exactmatchaccuracy_breakdown_GPT}
  \end{minipage}
  \vspace{-0.51cm}
\end{figure*}

\subsubsection{Accuracy for GSM8k and Trends Across Model Evolution}
\label{sec:trend}
In addition, we follow the same methodology as in previous work \cite{khattab2023dspy} for the evaluation of accuracy by focusing on the mathematical problem benchmark using a standard dataset, GSM8K~\cite{cobbe2021gsm8k}, with expected answers. This dataset allows us to perform an objective evaluation of accuracy across a variety of inputs. We sample 300 question-answer pairs and evaluate the accuracy of the programs implemented using \systemname{} and DSPy. Furthermore, we conduct the accuracy evaluation using 10 LLMs across generations throughout the model evolution to understand the accuracy trend as LLMs become more intelligent.

\textbf{\systemname{} vs. DSPy}. As shown in Figure \ref{fig:gsm_exactmatchaccuracy_openai_mtllm}, for the latest GPT models (GPT-4-turbo and GPT-4o) DSPy and \systemname{} perform similarly while \systemname{} outperform slightly for the most recent GPT-4o model. This shows that although \systemname{} hides all of the prompt engineering and type-conversion complexity, it achieves similar and sometimes even better performance accuracy through high quality automatic prompt synthesis and output interpretation. Even more interestingly, we observe that \emph{as the models continue to evolve and improve, the accuracy of programs implemented using \systemname{} improves significantly}, while DSPy's accuracy plateaued and also degraded a bit. \emph{This trend indicates that as models get better, the need for complex prompt engineering actually is largely reduced}. LLM's increasing capability to understand the traditional code, extract semantic meanings from the code and generate desirable output enabled \systemname{}'s high-accurate automatic prompt synthesis and output interpretation. 

\textbf{\systemname{} vs. DSPy (Compiled)}. It is also interesting to note that DSPy supports a compilation mode that requires additional training examples. It then searches across a set of prompt templates to optimize the prompt to improve accuracy, which could be time-consuming.  We observe that \systemname{} (with no training) outperforms DSPy (compiled with training) on all LLMs with the exception of GPT-4 (Figure~\ref{fig:gsm_exactmatchaccuracy_openai_mtllm}). These results further highlight the effectiveness of \systemname{} in automatically generating robust prompts using the semantic elements available in MT-IR.

We performed the same experiments with recent Llama models. Figure~\ref{fig:gsm_exactmatchaccuracy_llama_mtllm} shows a similar trend: \systemname{} accuracy improves as the Llama model gets better, indicating the decreasing need for complex prompt engineering for more intelligent models. Figure~\ref{fig:gsm_exactmatchaccuracy_breakdown_GPT} presents the failure / success breakdowns between the two approaches. It shows that for DSPy, the inaccuracy when using GPT models is often due to "type mismatch": e.g. the traditional code is expecting an integer output, yet the automatic type conversion in DSPy failed. For Llama, the degradation of DSPy's accuracy at the latest model is mostly due to longer execution time of the LLMs (timeout set to be 120s).

\subsection{Token Usage, Cost and Runtime}
\label{sec:token}

\begin{figure*}[t] 
  \begin{minipage}[c]{0.43\linewidth}
    \centering
    \includegraphics[width=\linewidth]{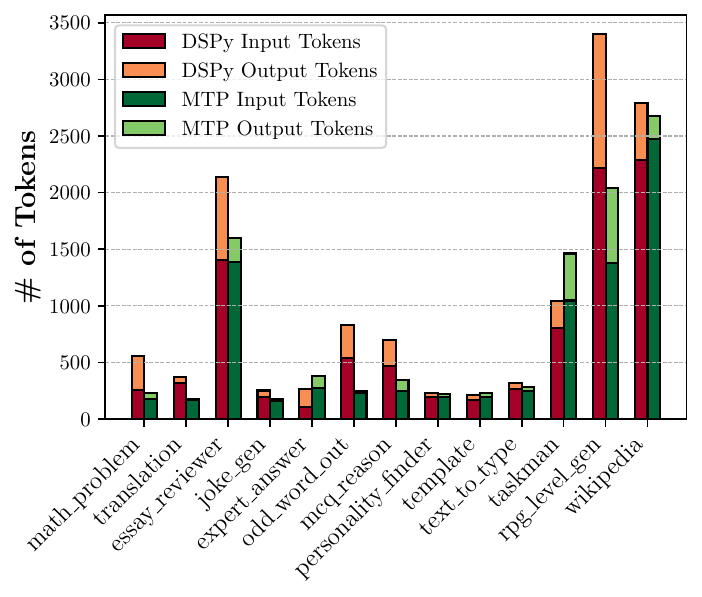}
    \caption{Token usage comparison between~\systemname{} and DSPy.}
    \label{fig:benchmark_token_usage}
  \end{minipage}
  \hfill 
  \begin{minipage}[c]{0.55\linewidth}
    \centering
    \includegraphics[width=\linewidth]{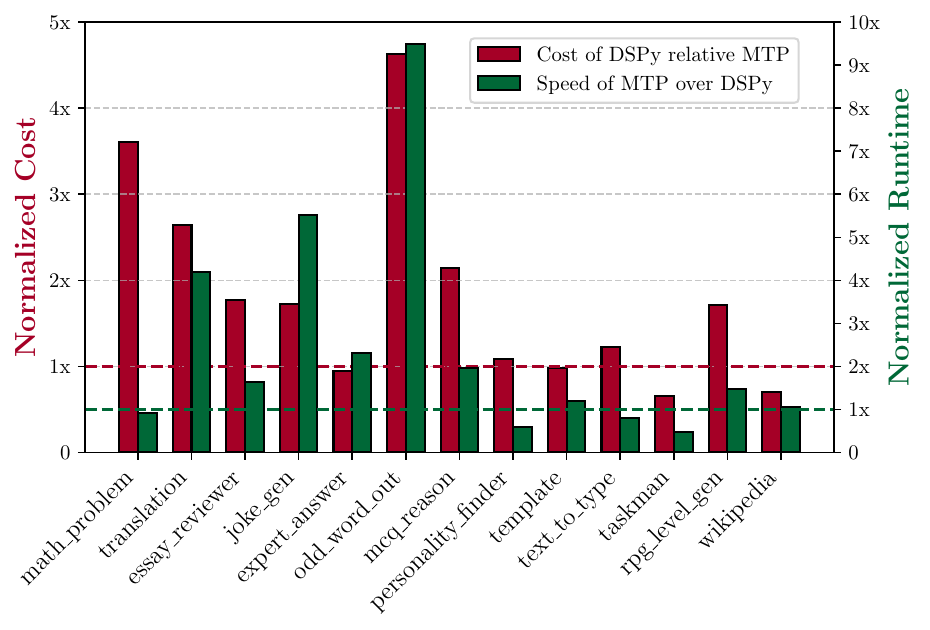}
    \caption{Cost and runtime speed comparison between~\systemname{} and DSPy.}
    \label{fig:benchmark_token_usage_runtime_cost}
  \end{minipage}
  \vspace{-0.3cm}
\end{figure*}

Token usage is an important metric since a higher token count leads to longer runtime (LLM inference time) and increased expenditure. Figure \ref{fig:benchmark_token_usage} presents the token usage comparison between DSPy (not-compiled) and \systemname{} across the benchmark suite both for prompt (input) and completion (output) token usage. Across the benchmark suite, we observe that \systemname{} consistently uses fewer tokens compared to DSPy, including the math problem using GSM8K dataset (first cluster of bars).
Figure \ref{fig:benchmark_token_usage_runtime_cost} presents the runtime improvement and cost reduction achieved by \systemname{} over DSPy. For each benchmark, the first bar shows the DSPy's cost relative to  \systemname{} (left y-axis) and the second bar shows the \systemname{}'s runtime improvement over DSPy (right-axis). The inference cost with the GPT-4o model is calculated using the OpenAI's pricing formula  \cite{openai_api_pricing}. The runtime is measured using cprofile. Figure \ref{fig:benchmark_token_usage_runtime_cost} demonstrates that \systemname{} achieves significant runtime improvement and cost reduction due to its more efficient token usage on majority of the benchmarks. As an example, for the \texttt{math\_problem} benchmark, we can see more that 3.5x cost savings while maintaining same speed as DSPy. On the other hand, for the \texttt{odd\_word\_out} benchmark we can observe more than 4x cost reduction as well as runtime improvement over DSPy.

\subsection{Sensitivity to Coding Practices}
\label{sec:sensitivity}

\begin{wrapfigure}{r}{0.4\textwidth}
    \centering
    \vspace{-0.58cm}
    \includegraphics[width=1\linewidth]{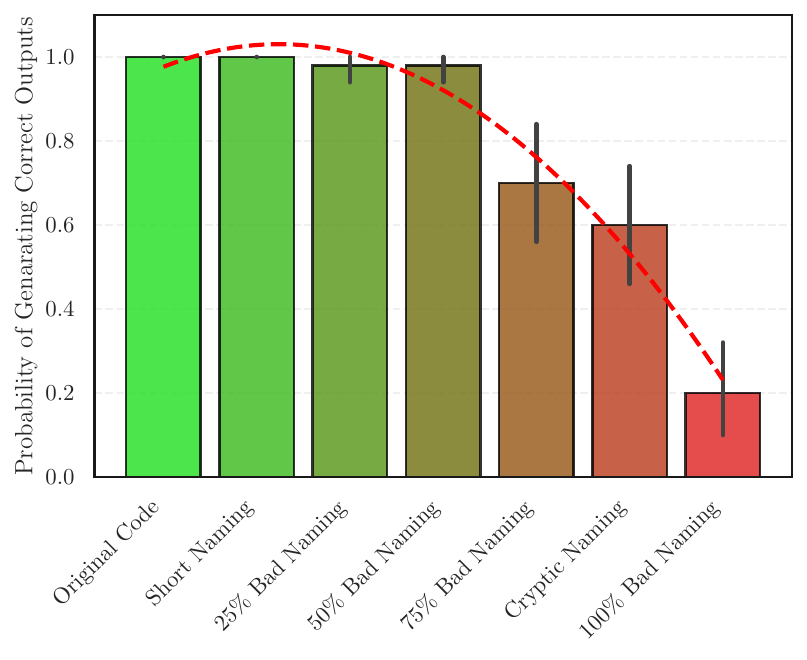}
    \caption{Sensitivity of \systemname{} to poor coding practices. The correct output probability declines slowly for first three experiments showing the resilience of \systemname.}
    \label{fig:sensitivity}
    \vspace{-0.5cm}
\end{wrapfigure}

MTP leverages semantics, enabling automated prompt generation for AI-Integrated applications. Here, we evaluate how sensitive \systemname{} is to poor coding practices, particularly when semantic richness is reduced.

To study this, we used the same video game level generation program from \S~\ref{sec:motivating_example}. Starting with the original implementation, we systematically introduced poor coding practices and measured the probability of generating a correct output out of 100 consecutive runs. Correctness of the output is defined in Table~\ref{tab:benchmarks}.

Figure~\ref{fig:sensitivity} shows a gradual decline in correctness of output as semantic quality deteriorates. We first examined the effects of shortened but still meaningful names, a common practice for maintainability. This had no impact on correctness. Next, we evaluated progressively worse naming conventions by replacing meaningful identifiers with increasingly abbreviated or single letter alternatives. We tested four levels of degradation: renaming 25\%, 50\%, 75\%, and 100\% of identifiers.

At 25\% and 50\% renaming, \systemname{} remained highly effective, with a correct output probability of 98\%. However, as renaming increased to 70\%, where nearly all data structures had poorly chosen names, correctness dropped to 70\%. When renaming reached 100\%, including function signatures, correctness declined sharply to 20\%, highlighting the limits of \systemname's resilience.

Additionally, we tested cryptic naming conventions, where identifiers retained internal consistency but lost intuitive meaning (e.g., renaming "Level" to "Boss" and "Wall" to "Chasm"). As shown in Figure~\ref{fig:sensitivity}, this resulted in a 60\% success rate, having a greater negative impact than 75\% abbreviation but still performing better than complete semantic loss.

These results demonstrate \systemname’s robustness, as the LLM can infer missing context, but excessive semantic degradation reduces accuracy. 

\begin{figure}[ht]
    \centering
    \begin{subfigure}[h]{0.48\linewidth}
        \centering
        \includegraphics[width=0.8\linewidth]{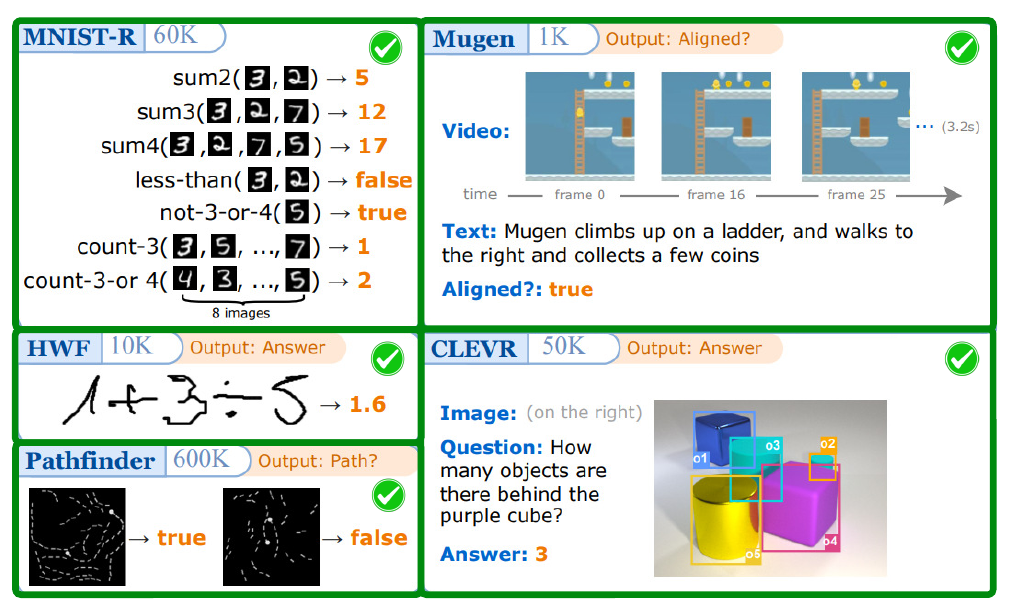}
        \caption{Multi-modal solutions from Scallop~\cite{li2023scallop}.}
        \label{fig:scallop-vision}
    \end{subfigure}
    \begin{subfigure}[h]{0.48\linewidth}
        \centering
        \begin{minipage}{\linewidth}
            \begin{python}
# CLEVR benchmark problem implemented in MTP

from mtp.core.llms import OpenAI
from mtp.core.types import Image

llm = OpenAI(model_name="gpt-4o")

can get_answer(img: Image, question: str) -> str by llm()

question:str = 'How many objects behind the purple cube?'
print(get_answer(Image('image.png'), question))
            \end{python}
        \end{minipage}
        \caption{Example CLEVR code snippet.}
        \label{fig:clevr-code}
    \end{subfigure}
    \caption{Implemented multi-modal solutions for visual tasks from Scallop (Left) and CLEVR code snippet (Right).}
    \Description{}
    \label{fig:scallop-mtt-vision}
\end{figure}

\subsection{\textbf{Using Multi-Modal LLMs to Solve Vision Problems}}
\label{app:multimodal}
There has been significant advancement recently in Multi-modal Generative AI models~\cite{wadekar2024MM-LLMevolution}.
These models, such as LLaVa~\cite{liu2023llava}, QWEN~\cite{yang2024qwen2technicalreport}, OpenAI's GPT-4o, Anthropic's Claude 3 and Google's Gemini can process both text and images and perform visual tasks such as object detection and visual reasoning.
Multi-modal LLMs achieve this by using specialized adapters that map different information modalities (text, images, video) into a unified vector space, enabling seamless reasoning across diverse input types.

\systemname{} leverages a similar approach by providing custom type interfaces that serve as adapters for different modalities. Just as multi-modal LLMs use internal adapters to handle various data types, developers can define custom interfaces such as \texttt{Image} or \texttt{Video} classes that seamlessly integrate with underlying system to interact with multi-modal LLMs. In this example, we demonstrate how ~\systemname{} generalizes leveraging such multi-modal generative models to build AI-Integrated programs using the same abstraction, reducing the developer effort at implementing such applications.

We select 5 vision benchmarks from related work Scallop~\cite{li2023scallop}. This set of problems represents a wide range of multi-modal tasks such as visual reasoning and visual question-answering, as shown in Figure~\ref{fig:scallop-vision}.
Using~\systemname{}, we are able to implement fully working solutions for these 5 vision problems using the same abstraction. The code snippet in Figure~\ref{fig:clevr-code} shows our implementation of one of the benchmark problems (CLEVR), which involves reasoning about an input image and providing an answer to a user's question.
As \systemname{} treats every problem using semantics and types, using images as inputs just comes down to defining the \texttt{Image} type used on line 4 in Figure~\ref{fig:clevr-code}. This adapter-like interface abstracts away the complexity of multi-modal model integration, demonstrating the wide range of use cases for \systemname{} while further reducing the need for developers to learn different frameworks for various AI-Integration problems.

Our evaluation confirms that MTP simplifies LLM integration by reducing developer effort (\ref{RQ1}) while maintaining accuracy (\ref{RQ2}). By minimizing code complexity and improving usability, MTP outperforms DSPy and LMQL without increasing token usage and runtime costs (\ref{RQ3}). Additionally, MTP shows a good degree of resilience to poor coding practices (\ref{RQ4}).

\section{Related Work}

Recent efforts to integrate Large Language Models (LLMs) into applications have emerged along two primary dimensions discussed in Section~\ref{sec:motivating_example}: AI-code generation and AI-integration. Works such as Pythoness~\cite{levin2025pythoness} directly address AI-code generation, while we focus on AI-integration in this paper.

Several frameworks have been developed to facilitate LLM integration into applications. DSPy~\cite{khattab2023dspy}, LMQL~\cite{Beurer2023LMQL}, and SGLang~\cite{zheng2023SGLang} introduce new libraries to help facilitating the integration of LLM into applications. While these frameworks provide additional tooling for LLM integration, they also introduce additional development complexity and new learning curves.

In addition to these frameworks there are tools built to streamline some parts of the prompt engineering process such as Outlines~\cite{willard2023outlines} and TypeChat~\cite{microsoft_typechat}. These tools focus on structured output generation using schemas and are introduced as complementary tools on top of prompt engineering which can improve accuracy, not as standalone frameworks that reduce prompt engineering effort. LangChain~\cite{langchain} and LlamaIndex~\cite{llamaindex} focus on simplifying building LLM application while integrating with external data sources and can be leveraged together with our work. With this approach, the developer can retain the novel language-level abstraction that automates away much of complexity while benefit from the additional functionality introduced in these libraries.

There has been extensive studies in specialized compilation approach for AI and ML models~\cite{ansel2024pytorch, chen2018tvmautomatedendtoendoptimizing, lattner2021mlir, 258858, rotem2018glow, davies2024journey, xia2024optimizing, hu2024optimal, pan2023recom}. In addition, recent works have also proposed novel runtime techniques and software system to accelerate training and inference of large scale models including large language models~\cite{orca-osdi, aminabadi2022deepspeed, 258953, 280768, guo2023olive, niu2021dnnfusion, zheng2022bytegnn, huang2021cosa, miao2024specinfer, oh2024exegpt, ahmad2024proteus, miao2024spotserve, kim2023dream, xu2024socflow, wang2024rap, chen2024magis}. Furthermore, researchers have explored architectural support for ML/AI models~\cite{zhong2024feasta, ghodrati2024tandem}. These work focus on accelerate training and inference of machine learning models. Our work focus on simplifying the development required to integrate LLM models in applications and can benefit from improved model performance.
\section{Conclusion}

Software is rapidly evolving from traditional logical code to AI-integrated applications that leverage generative AI and LLMs for application functionality. 
However, leveraging GenAI models in those applications is complicated and requires significant expertise and efforts. This paper presents meaning-typed programming (MTP),
a novel approach to simplify the creation of AI-integrated
applications by introducing new language-level abstractions
that hide the complexities of LLM integration. MTP automatically extracts the semantic meaning embedded in the traditional code, combined with dynamic information to dynamically synthesize prompts for LLMs and then convert the LLM output to the types that traditional code expects. We implemented MTP in a supersetted python language and demonstrate its effectiveness in reducing development complexity, improving runtime performance while maintaining high accuracy.


\begin{acks}
We would like to thank Chandra Irugalbandara for his contribution to the implementation of MTP.
\end{acks}

\section*{Data-Availability Statement}

The artifact accompanying this paper, which includes the implementation, benchmarks, and evaluation scripts, is available on \hyperlink{https://github.com/Jayanaka-98/mtllm-oopsla2025}{github.com/Jayanaka-98/mtllm-oopsla2025}~\cite{dantanarayana2025mtllm}.

\bibliographystyle{ACM-Reference-Format}
\bibliography{references}

\end{document}